\documentclass[12pt,letterpaper]{article}

\usepackage{graphicx,array}
\usepackage{url}
\usepackage{color}
\usepackage{latexsym}
\usepackage{amsthm}
\usepackage{amsmath}
\usepackage{amssymb}
\usepackage{amsfonts}
\usepackage[numbers,sort&compress]{natbib}
\usepackage{bm}
\usepackage{slashed}
\usepackage{mathrsfs}
\usepackage{enumerate}
\usepackage{tikz}
\usepackage{siunitx}
\usepackage{mdframed}
\usepackage{setspace}  
\usepackage{esvect}

\usepackage[utf8x]{inputenc}%
\usepackage{tcolorbox}%

\usepackage{hyperref} 
\hypersetup{
    colorlinks=true,       
    linkcolor=red,          
    citecolor=blue,        
    filecolor=magenta,      
    urlcolor=blue           
}
\usepackage[all]{hypcap} 

\usepackage{natbib}
\newcommand{\nc}{\newcommand}

\nc{\beq}{\begin{equation}}  
\nc{\eeq}{\end{equation}}  
\nc{\beqa}{\begin{eqnarray}}  
\nc{\eeqa}{\end{eqnarray}}  
\nc{\bit}{\begin{itemize}}  
\nc{\eit}{\end{itemize}}  

\usepackage{floatrow}
\newfloatcommand{capbtabbox}{table}[][\FBwidth]

\usepackage{blindtext}

\def\figureautorefname~#1\null{Fig.\,#1\null}
\def\tableautorefname~#1\null{Tab.\,#1\null}

\def\equationautorefname~#1\null{Eq.\,(#1)\null}

\title{ 
{\bf Dynamical Inflection Point Inflation}
\author{\large Yang Bai$^{\,\star}$ and Daniel Stolarski$\,^\diamond$}
\date{\small \it 
$^\star$Department of Physics, University of Wisconsin-Madison, Madison, WI 53706, USA\\
$^\diamond$Ottawa-Carleton Institute for Physics, Carleton University, 1125 Colonel By Drive, \\ Ottawa, Ontario K1S 5B6, Canada 
}
}

\begin{document}

\maketitle

\setlength{\parskip}{0.2ex}

\begin{abstract}	
We provide a mechanism to obtain multiple inflection points for the inflaton potential based on the Coleman-Weinberg potential. The key ingredient is the existence of zeros of the inflaton quartic coupling beta function, which can be simply realized by a sign change of a gauge coupling beta function at a massive threshold scale in gauge-Yukawa models. A universal inflaton potential emerges for a wide range of models and can accommodate the observational data in the small-field inflation scenario. While the ratio of tensor over scalar amplitudes is predicted to be small, the running of the scalar spectral index, $\alpha_s$, has branches with both positive and negative values. The positive branch with $\alpha_s \gtrsim 0.002$ will be tested by the coming cosmic microwave background observations. 
\end{abstract}

\thispagestyle{empty}  
\newpage  
  
\setcounter{page}{1}  


\newpage

\section{Introduction}\label{sec:Introduction}

The search for the underlying inflation models has been ongoing for around four decades~\cite{Guth:1980zm,Linde:1981mu}. With the impressively precise results from the cosmic microwave background (CMB) observations~\cite{Akrami:2018odb,Aiola:2020azj}, different inflation models have been favored or disfavored. For instance, single-field slow-roll models prefer a concave potential, $V''(\phi) < 0$ and a value of the tensor-to-scalar ratio $r$ below 0.1~\cite{Akrami:2018odb}. A class of traditionally well-motivated models are now shown to be in contradiction with current observational data (see \cite{Kallosh:2019jnl} for recent discussion). One simple attitude is to abandon those models and concentrate on other allowed, maybe less motivated, models. Another attitude is to see if any simple extensions or additional features of well-motivated models can be employed to make them viable. In this paper, we take the latter approach and focus on the Coleman-Weinberg (CW) inflaton potential, which is well-motivated in a wide range of field-theory models and has long been studied in the literature~\cite{Linde:1981mu,Guth:1982ec,Shafi:2006cs,Okada:2014lxa,Okada:2016ssd,Kallosh:2019jnl,Urbano:2019ohp}. 

For the simple CW potential containing the $\lambda\,\phi^4 \ln{\phi}$ term,
the inflationary data can only be fit with super-Planckian field excursions~\cite{Shafi:2006cs,Kallosh:2019jnl}. The large field scenario requires controlling an infinite tower of Planck suppressed operators, as well as possibly a knowledge of quantum gravity. Therefore in this work we focus on small field inflation. In that case, the CW potential predicts a spectral index $n_s$ of approximately $1 - 3/N_e$ with $N_e$ as the number of e-foldings. For $N_e$ around 60 to explain the horizon problem, $n_s$ is around 0.95, which is significantly smaller than the observationally measured value, $n_s =0.9691\pm 0.0041$~\cite{Akrami:2018odb,Aiola:2020azj}. In small field models, $n_s$ is controlled by the second derivative of the potential, and thus one needs to engineer a smaller value of $|V''(\phi)|$ to fit the data. 

This calculation of $n_s$ in the CW model is based on a leading-log approximation of the running of the coupling $\lambda$. If the running effects are important, one can compute a renormalization group improved potential 
\begin{equation}
V(\phi) \sim \lambda(\phi) \, \phi^4 ~,
\label{eq:schemPot}
\end{equation}
that can be utilized to reduce $|V''(\phi)|$ and increase the predicted $n_s$. To have $|V''(\phi)|\approx 0$ at some field point close to the initial inflation field value, one basically requires that the potential contains an inflection point. It is not new to apply an inflection point potential to inflation models~\cite{Allahverdi:2006iq,Baumann:2007np,Allahverdi:2008bt,Ballesteros:2015noa,Choi:2016eif,Dimopoulos:2017xox,Musoke:2017frr,Okada:2019yne}, but it is our purpose here to realize the inflection point potential based on a simple and dynamical field-theory setup without precise tuning of the parameters in the potential. 

The main point to realize the dynamical inflection point inflation based on the CW potential is the relation between the existence of inflection points away from the global minimum and zeros of the beta function of the quartic coupling $\beta_\lambda \equiv d\lambda/d\ln{\phi}$. Given the potential of Eq.~\eqref{eq:schemPot}, the first derivative of the inflaton potential is given by $V'(\phi) = \phi^3\,(4\lambda + \beta_\lambda)/4$, while the second derivative is $V''(\phi) \approx \phi^2\,(12 \lambda + 7 \beta_\lambda)/4$ after ignoring the suppressed $\beta'_\lambda$ term for a weakly-interacting theory. Inflection points of the potential are located at $\beta_\lambda \approx - 12\,\lambda/7$, and the extrema of the potential are located at $\beta_\lambda = -\lambda/4$. 

We can now see how the potential behaves starting from very large field values. The potential being bounded from below requires $\lambda(\phi) > 0$ for $\phi \rightarrow \infty$. As $\phi$ decreases, $\lambda$ deceases for $\beta_\lambda >0$ and the global minimum is located at a point in field space with $\lambda < 0$. Around the global minimum of the potential, one inflection point is anticipated with the field value slightly smaller than the minimum point value. To have an enough field traveling range for a viable inflation model, the inflection point is preferred to be parametrically separated from the global minimum point where the inflation ends. Additional inflection points can exist if the beta function has multiple zeros. As $\phi$ decreases, $\beta_\lambda$ passes one zero and becomes negative, which causes the coupling to increase. As $\beta_\lambda$ passes a second zero and becomes positive and if the coupling $\lambda$ stays  negative, one then has additional inflection points, which could be parametrically far away from the global minimum location. This behavior is explained in more detail in Section~\ref{sec:CW} and shown qualitatively in Fig.~\ref{fig:lambda}.

Although the above description seems to be involved, it is very simple to have multiple zeros for the beta function, especially when there are multiple couplings entering $\beta_\lambda$. In this paper, we will choose one example model with a non-Abelian gauge symmetry and fermions to illustrate our purpose. The leading contributions to $\beta_\lambda$ come from the gauge coupling $\propto g^4$ and the Yukawa coupling $\propto - y^4$. Ignoring the running of $y$, the running of $g$ can cause it to have a minimum value at some scale $M$, where $\beta_g$ changes sign by decoupling some massive states. For a proper choice of the minimum value of $g$, one can realize two zeros for $\beta_\lambda$ and hence the dynamical inflection point inflation. 

To fit the inflation observational data, we will also parametrize the inflation potential based on a wide range of possible gauge-Yukawa models with different gauge groups and matter content. This phenomenological potential given in Eq.~\eqref{eq:pheno} can be directly compared to CMB data in future studies. We will show that the observed value of the spectral index $n_s$ can be easily accommodated for the generalized CW potential with non-trivial coupling running. Inflation begins at a field point close to the inflection point and 60 e-foldings of inflation to solve the horizon problem can be easily accommodated. Tensor modes are expected to be small, but the running of the scalar spectral index is predicted as a function of one parameter with compact range and one discrete parameter. 

To proceed further, we also want to compare our study to some closely related works in Refs.~\cite{Ballesteros:2015noa,Dimopoulos:2017xox,Okada:2019yne} that also use a potential similar to Eq.~\eqref{eq:schemPot}. In~\cite{Ballesteros:2015noa}, two loop corrections are used to modify the potential, but it is a large field inflation model. In~\cite{Dimopoulos:2017xox}, the potential is modified by Planck suppressed higher dimensional operators, and in the potential of \cite{Okada:2019yne}, the CW plateau is the global minimum while inflation occurs near a single zero of $\beta_\lambda$ that is parametrically displaced from plateau. Both~\cite{Dimopoulos:2017xox,Okada:2019yne} require a precise tuning of particle physics parameters. A closely related reference to our study is Ref.~\cite{Okada:2016ssd}, where the $U(1)_{B-L}$ Higgs field is used as the inflaton to achieve a large-field dynamical inflection point inflection. In this work, we present a small field model where the CW plateau is utilized for inflation and, while some parameters have to be roughly the same size as one another, no precise tuning is needed. 
 
Our paper is organized as follows. In Section~\ref{sec:slow-roll}, we give a brief review of the standard results for slow roll inflation. In Section~\ref{sec:CW}, we discuss how to realize the inflection points in the CW potential, while in Section~\ref{sec:realization} we provide an explicit field-theory realization. We then fit to the inflation observables in Section~\ref{sec:fit}. In Section~\ref{sec:attractor} we check whether the inflationary solution in this potential is an attractor and show that the slow roll analysis is justified, and we conclude in Section~\ref{sec:conclusion}. In Appendix~\ref{app:colourflavour}, we present formulae for models with a more general color and flavor structure. 

\section{Slow roll inflation}
\label{sec:slow-roll}

Given an inflaton potential $V(\phi)$, we can define the slow roll parameters~\cite{Baumann:2009ds}
\beqa
\setlength{\jot}{5pt}
\begin{aligned}
\epsilon_{\rm v} &= \frac{1}{2} \left. \left( \frac{V'(\phi)}{V(\phi)} \right)^2 \right\vert_{\phi = \phi_{\rm i}} ~,\\
\eta_{\rm v} &=\left. \frac{V''(\phi)}{V(\phi)}  \right\vert_{\phi = \phi_{\rm i}}  ~,\\ 
\xi_{\rm v}^2 &=\left. \frac{V'''(\phi)V'(\phi)}{V(\phi)^2}  \right\vert_{\phi = \phi_{\rm i}}  \label{eq:xi}~,
\end{aligned}
\eeqa
where $\phi_{\rm i}$ 
is the point in field space where the cosmological scales leave the horizon.
We are using the reduced Planck scale $M_{\rm Pl} = 1$ units with $M_{\rm Pl} = 1/\sqrt{8\pi G_{\rm N}} \approx 2.4 \times 10^{18}$ GeV and $G_{\rm N}$ as the Newton constant. The magnitude of these parameters must be $\ll 1$ to be within the slow-roll regime. In the slow-roll approximation, these parameters of the potential can be mapped onto observables that can be measured in the CMB. For the scalar and tensor components of perturbations, the spectra can be approximated by power laws with $\mathcal{P}_{\mathcal{R}} = A_s(k/k_*)^{n_s - 1 +\frac{\alpha_s}{2}\,\ln{k/k_*}}$ and $\mathcal{P}_t = A_t (k/k_*)^{n_t}$, respectively. Here, $n_s$ is the scalar spectral index; $\alpha_s \equiv dn_s/d\ln k$ is the running of the scalar spectral index; $n_t$ is the tensor spectral index; $k_*=0.05\,\mbox{Mpc}^{-1}$ is a pivot scale~\cite{Akrami:2018odb}; the ratio of the tensor over scalar amplitude is $r \equiv A_t /A_s$; the tensor spectral index is $n_t \simeq -r/8$ for single field slow-roll inflation. The mapping to the potential properties is given by 
\beqa
n_s &\approx& 1-6\,\epsilon_{\rm v} + 2\,\eta_{\rm v} ~,\label{eq:ns}\\
r&\approx& 16\,\epsilon_{\rm v} ~, \\
\alpha_s &\approx& 16\,\epsilon_{\rm v}\,\eta_{\rm v}  - 24\, \epsilon_{\rm v}^2 -2 \, \xi_{\rm v}^2 \label{eq:alpha} ~, \\
A_s &\approx& \frac{1}{12\pi^2}\frac{V^3}{V'^2} ~.
\eeqa
Note that the first three observables are invariant under an overall rescaling of the inflation potential $V\rightarrow c V$, but not the fourth one. 

On the observational side, the combination of the Planck~\cite{Akrami:2018odb} and Atacama Cosmology Telescope (ACT)~\cite{Aiola:2020azj} has 
\beqa
\label{eq:exp-ns}
n_s &=& 0.9691\pm 0.0041 ~, \\
\label{eq:exp-r}
r &<& 0.067   ~, \\
\label{eq:exp-alphas}
\alpha_s &=& 0.0023\pm 0.0063 ~,\\
\label{eq:exp-As}
A_s &=& (2.189 \pm 0.053)\times 10^{-9} ~, 
\eeqa
with $68\%$ confidence level or $95\%$ upper limit. For simplicity, we will ignore correlations between the parameters. The upcoming Simons Observatory~\cite{Ade:2018sbj} will reduce the error on $\alpha_s$ by a factor of 10, and it will be sensitive to $r > 1.4\times 10^{-3}$.

Inflation ends when the slow-roll conditions are broken or  $V'/V \simeq 1$, at the point $\phi_{\rm e}$ in field space. The number of $e$-foldings before the inflation ends can be approximated by
\beqa
N_e \approx \int_{\phi_{\rm e}}^{\phi_{\rm i}} \frac{V(\phi)}{V'(\phi)}\, d\phi ~.
\label{eq:ne}
\eeqa
To explain the horizon problem, the required number of $e$-foldings has an upper bound of around 65 for the simple early-universe evolution after the end of inflation till Big Bang Nucleosynthesis (BBN), but can be much larger for a low reheating temperature or a non-standard cosmology~\cite{Dodelson:2003vq,Liddle:2003as}. In our study here, we will try to accommodate $N_e$ in the range of $50$--$60$, but not try to fit a specific value.

\section{Inflection points in the Coleman-Weinberg potential}
\label{sec:CW}

The dynamical inflection point inflation in our model is based on the well-studied CW Potential~\cite{Coleman:1973jx}, which has been adopted to realize inflation in the literature~\cite{Linde:1981mu,Guth:1982ec,Shafi:2006cs,Okada:2014lxa,Okada:2016ssd,Kallosh:2019jnl,Urbano:2019ohp}. The CW potential has the form of $V(\phi) =  \lambda \left\{ \phi^4\left[ \ln(\phi/f )-\frac{1}{4} \right] + f^4/4\right\}$, where $\phi=f$ is the minimum of the potential and the last term adjusts the cosmological constant to zero. This potential is attractive because it is very flat for $\phi \ll f$ easily allowing slow roll conditions to be satisfied. 

To fit the observed data in Eqs.~\eqref{eq:exp-ns}-\eqref{eq:exp-As} and $N_e \sim 60$, both large and small initial field values have been explored. For the large field value with $\phi_{\rm i} \approx 23\,M_{\rm Pl}$, one can accommodate the observed value of $n_s$ in Eq.~\eqref{eq:exp-ns} and satisfy the constraint of $r$ in Eq.~\eqref{eq:exp-r} for $N_e\sim 60$ (see Refs.~\cite{Shafi:2006cs,Urbano:2019ohp} for instance). Given the trans-Planckian field excursions, the UV-completion of the model, possibly including quantum gravity, is needed. On the contrary, the small field inflation does not suffer the potential trans-Planckian problem, although a good fit to $n_s$ cannot be obtained (at least for $\lambda$ independent of $\phi$). The situation is similar to other ``hilltop" potential with $V=V_0( 1 - \phi^4/m^4 +\cdots)$ in the limit of $m \ll M_{\rm Pl}$. The predicted spectral index is $n_s = 1 - 3/ N_e$ and is 0.95(0.94) for $N_e = 60(50)$, which is disfavored by the observed value in Eq.~\eqref{eq:exp-ns} (see Ref.~\cite{Kallosh:2019jnl} for recent discussion). This potential can be accommodated with $N_e \sim 90$, which can fit the data with non-standard cosmological histories after inflation~\cite{Dodelson:2003vq,Liddle:2003as}.

The simple expression for $N_e$ that is independent of $f$ can be understood as follows. For $f\lesssim M_{\rm Pl}$, the number of e-foldings is approximately given by $N_e \sim f^4/\phi_{\rm i}^2$, ignoring $\mathcal{O}(1)$ constants and logarithms. We then get that $\epsilon_{\rm v} \sim f^4/N_e^3$ and $\eta_{\rm v} \sim 1/N_e$. So we see, it is generic in these types of potentials for sub-Planckian field excursions, that $\epsilon_{\rm v} \ll \eta_{\rm v}$ and that $n_s$ is controlled by $\eta_{\rm v}$ and thus the second derivative of the potential at $\phi_{\rm i}$. Since the prediction of the CW potential is that $n_s$ is too small, we need to reduce the magnitude of the second derivative. This can be done if $\phi_{\rm i}$ is near an inflection point $\phi_{\rm inf}$ where $V''(\phi_{\rm inf}) = 0$~\cite{Ballesteros:2015noa,Dimopoulos:2017xox,Okada:2019yne}.

To realize a viable small-field inflation potential with an inflection point, we generalize the CW potential by considering the Renormalization Group (RG) improved version.  Taking into account the running of the quartic coupling, one has 
\beqa
V(\phi) &=& \frac{1}{4}\,\lambda(\phi) \,\phi^4 + V_0 ~,
\label{eq:genCW}
\eeqa
where the coupling $\lambda$ is a function of the renormalization scale that is chosen to be the inflaton field value, $\phi$, and $V_0$ is a constant such that the potential is approximately zero at the global minimum.  Different from the normal CW potential, we will take $\lambda(\phi)$ as determined from its full one-loop RG running rather than the leading log term.

Before we present an explicit model and expression for $\lambda(\phi)$, we first discuss how the general potential in Eq.~\eqref{eq:genCW} can contain an inflection point, where $V''(\phi_{\rm inf}) = 0$ (see also Ref.~\cite{Okada:2014lxa} for similar discussion). For $\phi_{\rm i}$ close to $\phi_{\rm inf}$, the absolute value of $\eta_{\rm v}$ is thus suppressed and the spectral index $n_s$ can be increased to fit the observed value.  The first and second derivatives of the above potential are given by
\beqa
\label{eq:potential-first-deriv}
V'(\phi) &=& \left(  \lambda + \frac{1}{4}\beta_\lambda \right)\phi^3 ~,\\
\label{eq:potential-second-deriv}
V''(\phi) &=& \frac{1}{4}\left( 12 \lambda + 7\beta_\lambda + \beta'_\lambda\right)\phi^2 ~,
\eeqa
with $\beta_\lambda \equiv d \lambda/d\ln\phi$ and $\beta'_\lambda \equiv d \beta_\lambda/d\ln\phi$. If the perturbative loop expansion is valid, then $\beta_\lambda$ is parametrically one-loop and $\beta'_\lambda$ is of two-loop size and can be ignored in Eq.~\eqref{eq:potential-second-deriv}. In our explicit examples below, the $\beta$ function will be accidentally small and the $\beta'$ term will be numerically important. In this section we will use the simplifying assumption that $\beta'$ can be ignored so that these equations can be solved analytically and we can get a rough sense of their parametric behavior. In that case, extrema of the potential therefore satisfy
\begin{equation}
\lambda(\phi_{\rm ext} ) = -\frac{1}{4}\,\beta_\lambda(\phi_{\rm ext} ) ~,
\label{eq:min}
\end{equation}
and inflection points $\phi_{\rm inf}$ occur when 
\begin{equation}
\lambda(\phi_{\rm inf}) \approx -\frac{7}{12} \beta_\lambda(\phi_{\rm inf}) ~.
\end{equation}

In a theory with only a scalar inflaton field, $\beta_\lambda \sim \kappa\,\lambda^2$ with  $\kappa = (16\pi^2)^{-1}$ being the usual loop factor, and Eq.~\eqref{eq:min} cannot be satisfied in the perturbative regime. In a more complicated model with other couplings of the inflaton scalar that can be schematically represented as $g$, then we have $\beta_\lambda \sim \kappa\left(g^4 + g^2 \lambda +  \lambda^2 \right)$, where for the remainder of this section we ignore order one factors, but we provide a more complete calculation in the next section. In this scenario, a minimum of the potential and an inflection point can both be found with $\lambda \sim -\kappa\,g^4$, allowing us to ignore the terms proportional to $\lambda$ in $\beta_\lambda$. A more careful calculation (neglecting the running of other couplings, $g$) shows that the ratio of the field point of the inflection point to that of the minimum is $\phi_{\rm inf}/\phi_{\rm min} \approx e^{-1/3} \approx 0.7$. These are too close in field space to achieve 60 e-foldings with sub-Planckian field excursions.

\begin{figure}[t!]
	\label{fig:lambda}
	\begin{center}
		\includegraphics[width=0.6\textwidth]{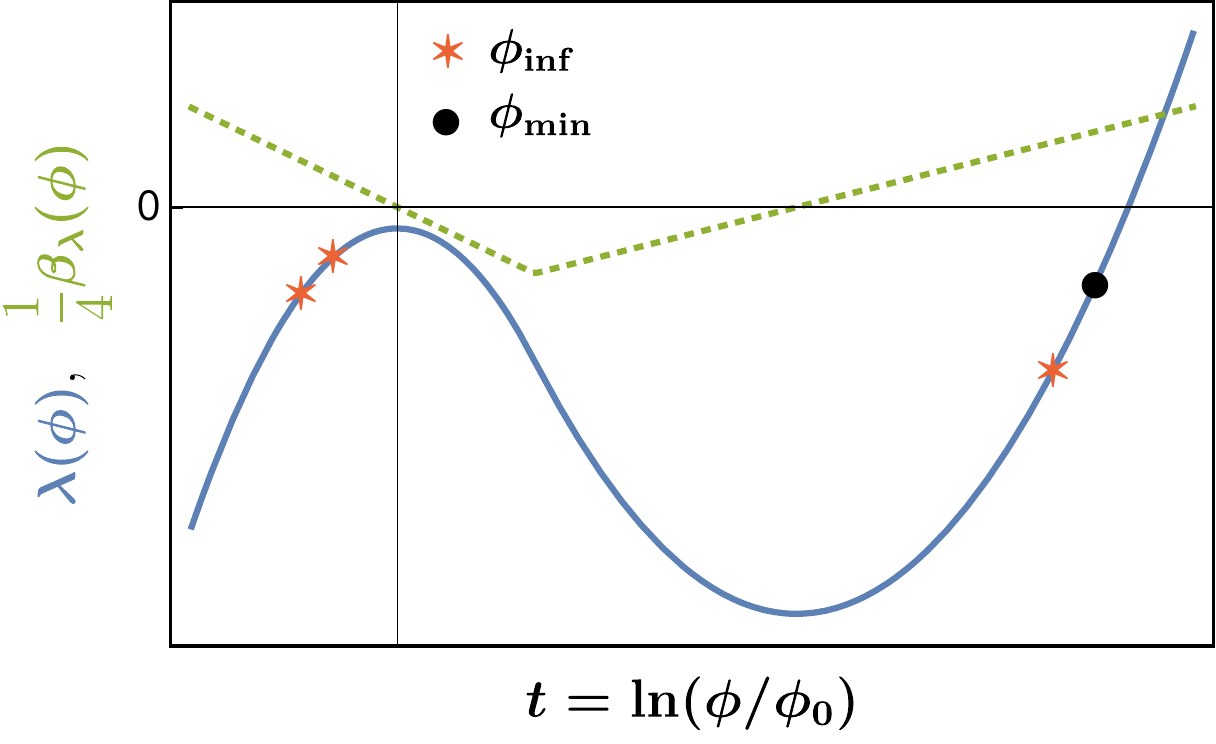} 
		\caption{Schematic plot for the quartic coupling $\lambda(\phi)$ (solid, blue) as a function of renormalization scale or $t \equiv \ln{(\phi/\phi_0)}$. We also show $\beta_\lambda/4$ (dotted, green), which, via Eq.~\eqref{eq:min}, can be used to find the location of the global minimum of the inflaton potential in Eq.~\eqref{eq:genCW} which is labelled by the solid black circle. The red stars show the locations of the inflection points. One inflection point is close to the global minimum of the potential, while the other two inflection points are further away and located at smaller values of $\phi$. }
	\end{center}
\end{figure}

This problem can be solved if there are two locations in field space where $\beta_\lambda = 0$ that are parametrically separated. In Fig.~\ref{fig:lambda}, we present a schematic plot to illustrate this general point. The two $\beta_\lambda = 0$ points correspond to the maximum $\lambda_{\rm max}$ and minimum $\lambda_{\rm min}$ of $\lambda$ as a function of $t \equiv \ln{(\phi/\phi_0)}$. On the right side of $\lambda_{\rm min}$, the black dot point shows the location of the global minimum of the potential where Eq.~\eqref{eq:potential-first-deriv} is satisfied, and the red star point shows the inflection point that is very close to the global minimum point of the potential. On the left side of the $\lambda$ maximum value, there are two more inflection points with one near $\lambda_{\rm max}$. Their field value $\phi_{\rm inf}$ can be exponentially suppressed compared to $\phi_{\rm min}$ because of the RG running effects. Therefore, the small-field inflation could be implemented in a simple way. If the scale $\phi_{\rm i}$ where cosmological scales exit the horizon is near one of the two inflection points on the left side, one can achieve $\eta_{\rm v} \approx -0.016$ to get the correct scalar spectral index.

In order for this to be a viable inflation potential, the inflaton must slowly roll from the initial point down to the minimum, which means that the first derivative of the potential cannot change sign, because in that case, the potential will have a local minimum and the inflaton could get trapped. This is achievable because specifying that $\beta_\lambda$ has two zeros leaves an overall shift in $\lambda$ unspecified. Thus we can choose the a boundary condition, $\lambda_0 = \lambda_{\rm max}$ sufficiently small such that near the second zero of $\beta_\lambda$, Eq.~\eqref{eq:potential-first-deriv} cannot be satisfied and  $dV/d\ln\phi$
does not change sign. Therefore, the inflaton can smoothly roll all the way to the minimum of the potential. 

\section{Field-theory realization}
\label{sec:realization}

We now give a field theoretic construction of the scenario described above with an inflation potential given Eq.~\eqref{eq:genCW} such that the $\beta$ function of the quartic coupling $\lambda$ has two zeros. 
We begin with an $SU(N_c)$ gauge theory with gauge coupling $g$, $N_{\rm UV}$ vectorlike fundamental flavors of fermions $\psi$, and a scalar inflaton $\phi$ charged under the fundamental of the gauge group. We require $N_{\rm UV} > (22\,N_c - 1) / 4$ or $\beta_{g} > 0$ such that the gauge coupling grows at higher energies (the Landau-pole scale is generically much higher than the Planck scale). At the scale $M$, some of the fundamental fermions acquire a mass, leaving $N_{\rm IR}$ light fermions, and $N_{\rm IR} < (22\,N_c - 1) / 4$ such that the gauge coupling grows towards the IR at scales below $M$. We also add $N_{\rm S}$ vectorlike gauge singlets $\chi$ allowing us to write down a Yukawa coupling $\mathbf{Y}$ between the fermions and the inflaton. Thus the Lagrangian is given as follows:
\begin{eqnarray}
{\cal L} = &-&\frac{1}{4}G^{a\,\mu\nu}G^a_{\mu\nu} \,+\, (D_\mu \Phi)^\dagger (D^\mu \Phi) 
 \,+\, i \, \bar{\chi}\slashed{\partial} \chi \,+\,  i\, \bar{\psi} \slashed{D} \psi \,-\, \sum_{i=N_{\rm IR}+1}^{N_{\rm UV}} M\, \bar{\psi}_i\psi_i \nonumber\\
 &-& \frac{\lambda}{4} \left(\Phi^\dagger \Phi \right)^2 - \left( \Phi\, \bar{\psi} \,  \mathbf{Y}\, \chi + {\rm h.c.}\right) ~,
 \label{eq:lagrangian}
\end{eqnarray}
where $\Phi^\dagger \Phi \equiv \phi^2/2$ with the radial field $\phi$ as the inflaton field. $\mathbf{Y}$ is an $N_{\rm UV} \times N_{\rm S}$ matrix at scales above $M$ and an $N_{\rm IR} \times N_{\rm S}$ at lower scales when we consider RG running equations. Thus we have a tree-level potential for the scalar simply given by quartic term proportional to $\lambda$. 

For simplicity, we do not include mass terms for the singlet, $\chi$, for the light charged fermions $\psi_i$, $i\leq N_{\rm IR}$, and for the scalar $\Phi$. The final omission is a requirement for the CW mechanism. All masses could be present in the theory as long as those states are parametrically lighter than the relevant mass scales. For example, for the scalar, we simply require $m_\phi^2 \ll \lambda \phi^2$. This hierarchy of masses is put in by hand to get a suitable inflation potential.


For simplicity, we first specify $N_c=2$, $N_{\rm IR} = N_{\rm S}=1$. For this case, only the light flavor has a Yukawa coupling denoted $y$ and taken to be real. As we will see below, this case has sufficient freedom to generate the general inflaton potential, but we describe a more general color and flavor structure in Appendix~\ref{app:colourflavour}. We now compute the RG improved loop corrections with the one-loop $\beta$ functions for the three coupling constants: 
\begin{eqnarray}
\beta_g &=& -\kappa\, \left(\frac{43}{6} - \frac{2}{3}\,n_f\right)\, g^3 \label{eq:betag}  ~,\\
\beta_{y} &=& \kappa
\left(\frac{5}{2} \, y^3
-\frac{9}{4} \,g^2\, y \right)\label{eq:betay}  ~, \\
\beta_\lambda &=& \kappa \left(\frac{9}{8}\, g^4 - 2\,y^4
- 9 \,g^2\, \lambda + 4\, y^2 \,\lambda + 24\, \lambda^2 \label{eq:betal}
\right)  ~,
\end{eqnarray}
where $\kappa = (16\pi^2)^{-1}$ and $n_f$ is the number of flavors kinematically accessible with $n_f = N_{\rm UV}$ for $\phi > M$ and $N_{\rm IR}=1$ for $\phi < M$. If we work in the regime where $y,g \gg \lambda$, then we see that $\beta_{\lambda} \approx 0$ if 
\begin{equation}
y=\frac{\sqrt{3}}{2}\,g ~.
\label{eq:yboudnary}
\end{equation} 
From our construction, we can see that $g$ grows both towards the UV and IR from the scale $M$ where some of the fermions are integrated out. So, the minimum value of $g$ happens at the scale $M$. Therefore, if $y(M) > \sqrt{3}\,g(M) /2$ and $g$ runs faster than $y$, this setup achieves our desired goal of having two zeros for $\beta_\lambda$ that are parametrically separated from one another.

Working in the parameter space where $y(M) > \sqrt{3}\,g(M) /2$, we can choose a reference scale $\phi_0$ to be one of the points in field space where Eq.~\eqref{eq:yboudnary} is satisfied and $\beta_\lambda$ vanishes. There are two such points, so we will choose $\phi_0 < M$ (or the location at the local maximum value of $\lambda$ in Fig.~\ref{fig:lambda}). As this potential is classically scale invariant, $\phi_0$ is the only scale and thus sets the overall energy of the potential. All other relevant scales, such as $\phi_{\rm i}$ will be proportional to $\phi_0$ by dimensional analysis. The other point where Eq.~\eqref{eq:yboudnary} is satisfied will be near the absolute minimum of the potential. 

With this choice of $\phi_0$ we can approximately solve the differential equations for the couplings as a function of scale. 
If we define $t\equiv \ln(\phi/\phi_0)$ and $t_M \equiv \ln(M/\phi_0) > 0$, then the solutions for the gauge and Yukawa couplings to leading order in the loop approximation ($\kappa \ll 1$) is given by:
\begin{eqnarray}
g(t) &=& g_0 - \kappa \,g_0^3 \left[\frac{13}{2}\, t - \frac{2}{3}\,(N_{\rm UV}-1)(t-t_M)\Theta(t-t_M) \right]  ~, 
 \label{eq:gysol-g}
\\
y(t)&=& y_0 - \frac{1}{2}\kappa\, y_0^3 \, t  ~,
 \label{eq:gysol}
\end{eqnarray}
 with $g_0$, $y_0$  being the couplings at the scale $\phi_0$, and $y_0$ and $g_0$ chosen such that Eq.~\eqref{eq:yboudnary} is satisfied. The function $\Theta(x)$ is the Heaviside step function which appears as a leading order approximation of massive states contributing to the renormalization group running only at scales above their masses.
 The running of $g$ is therefore faster than the running of $y$ for $\phi < M$ partially because of a cancellation of the two terms in Eq.~\eqref{eq:betay}. The running is also faster for $\phi > M$ as long as $N_{\rm UV} > 11.2$, and by construction we already have $N_{\rm UV} > 10.75$, so this is only a mild additional constraint. 
 
Ignoring terms proportional to $\lambda$ in Eq.~\eqref{eq:betal}, we can integrate the $\beta_\lambda$ to solve for the quartic coupling as a function of scale:
 \begin{equation}
 \lambda(t) = \lambda_0 \,-\, \frac{441}{32}\,\kappa^2\,g_0^6\, t^2 \,+\,\frac{3}{2} \,\kappa^2\, g_0^6\, (N_{\rm UV}-1)(t\,-\, t_M)^2\,\Theta(t-t_M) ~.
\label{eq:lsol}
 \end{equation}
We have used \eqref{eq:yboudnary} to eliminate $y_0$. 
If $\lambda_0 \sim \kappa^2 g_0^6$, then ignoring the $\lambda$ terms in $\beta_\lambda$ is a good approximation. We have also chosen $N_{\rm UV}$ to be large enough that at large $t$, $\lambda(t)$ increases and the inflaton potential is bounded from below.

\begin{figure}[t!]
	\label{fig:potential-example}
	\begin{center}
		\includegraphics[width=0.46\textwidth]{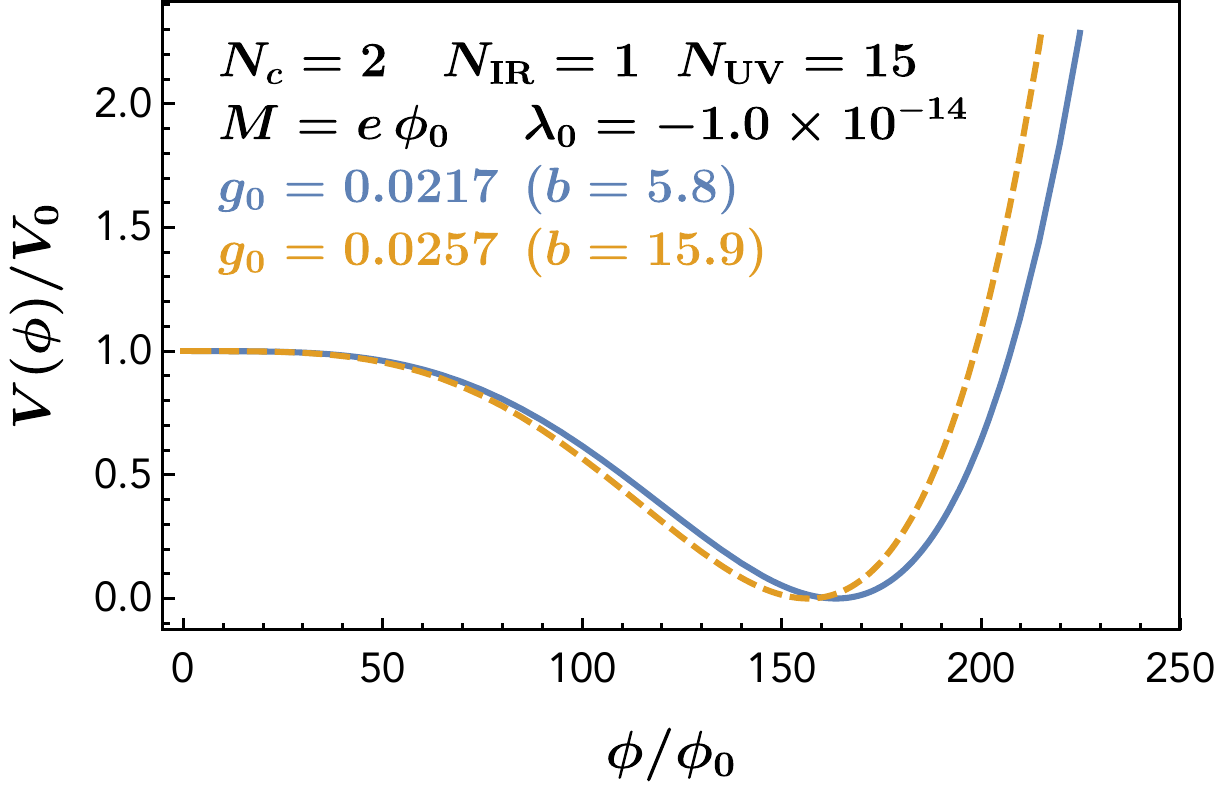} \hspace{3mm}
	         \includegraphics[width=0.46\textwidth]{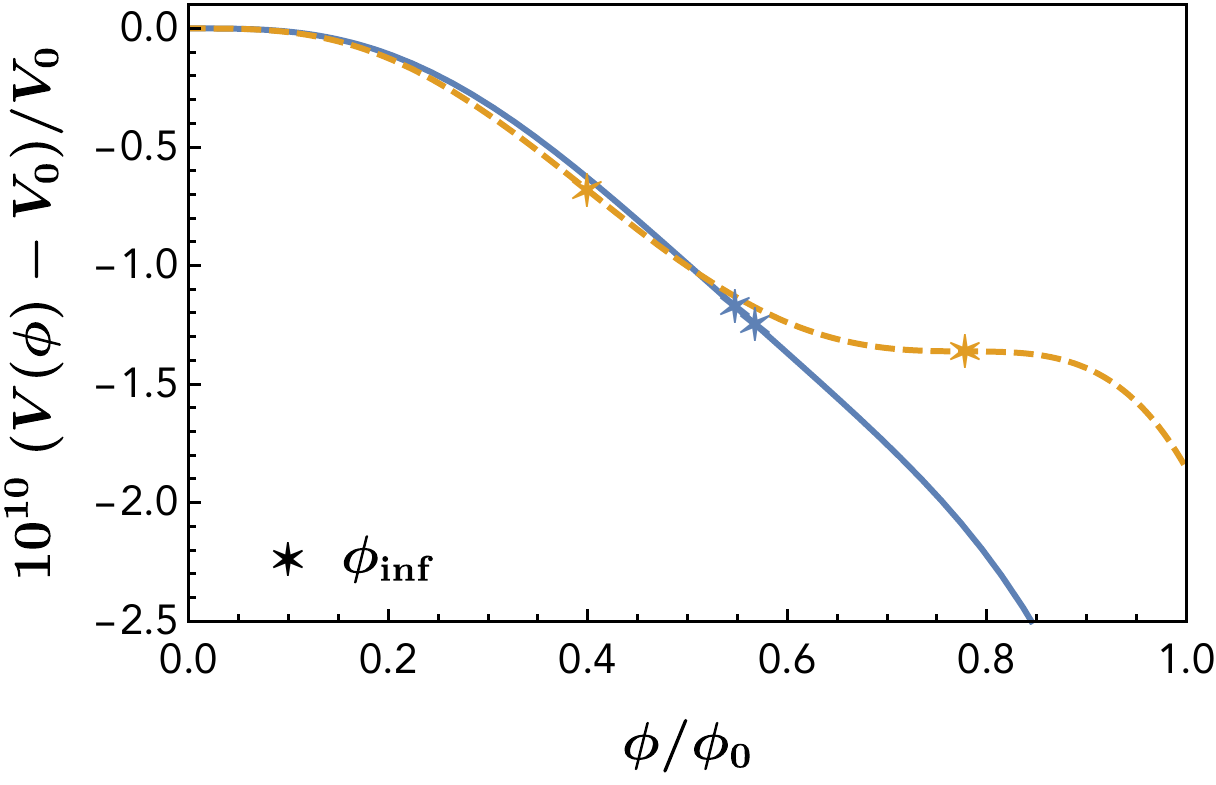} 
		\caption{Two example potentials based on Eq.~\eqref{eq:genPot} or Eq.~\eqref{eq:pheno}. The left panel has the general features similar to the ordinary CW potential, while the right panel zooms in on the small $\phi$ field region.  The inflection points are labeled by the star points. Here, $M= e\, \phi_0$ with $e = 2.718$. The parameter $b$ will be defined later in Eq.~\eqref{eq:pheno}. }
	\end{center}
\end{figure}

The potential for the inflation can be found by plugging $\lambda[\ln(\phi/\phi_0)]$ from Eq.~\eqref{eq:lsol} into Eq.~\eqref{eq:genCW}:
\begin{equation}
V(\phi) =   
\frac{\lambda_0}{4} \, \phi^4 \,
 \left[ 1
- \frac{441\,\kappa^2\,g_0^6}{32\,\lambda_0} \ln^2\left(\frac{\phi}{\phi_0} \right)
+\frac{3\,\kappa^2\,g_0^6}{2\,\lambda_0} \left(N_{\rm UV}-1\right)\ln^2\left(\frac{\phi}{M}\right) \,\Theta(\phi-M) 
\right]
+  V_0 ~.
\label{eq:genPot}
\end{equation}
For this potential to be viable requires $\lambda_0 < 0$, because then for field values $\phi < M$ the potential decreases, while for field values $\phi \gg M$ the potential increases. For $\phi \ll \phi_0$ the potential is very flat, much like the original CW potential. Thus, as long as $\phi_0$ is not too close to $M$, the global minimum of the potential will occur at some field value $\phi_{\rm min} > M$. The value of $V_0$ can thus be set by the requirement that the cosmological constant is very small at the minimum, $V(\phi_{\rm min})\approx0$. In Fig.~\ref{fig:potential-example}, we show two example potentials with inflection points in the small $\phi$ region. We see that the full potential looks very similar to the original CW potential, but that the behavior on the plateau is non-trivial, containing inflection points. 

\section{Fit to inflation observables}
\label{sec:fit}

In order to get a viable slow-roll potential model, we need the inflaton to roll towards the absolute minimum,\footnote{We do not consider the possibility that the inflaton settles in a cosmologically long-lived local minimum. This does not arise in the types of potential we consider here.} thus, as discussed above, $\lambda_0 < 0$ so that the derivative of the potential at $\phi_0$ is negative. Therefore, we can use the following phenomenological potential to parameterize dynamical inflation point inflation:
 \begin{equation}
 V(\phi) = -\frac{a}{4}\,\phi^4\,\left[ 1 + b\, \ln^2\left(\frac{\phi}{\phi_0} \right)- c \, \ln^2\left(\frac{\phi}{M}\right)\,\Theta(\phi-M) \right]\,+\,  a\,V_0 ~,
 \label{eq:pheno}
 \end{equation} 
 with $a>0$, $c>b>0$, and $M>\phi_0$. The requirement that the potential at the minimum is vanishing also imposes $V_0 > 0$. Note that we have rescaled $V_0$ by $a$ for convenience.  This phenomenological potential contains the same number of free parameters as the field theory potential of Eq.~\eqref{eq:genPot}, justifying the simplifying assumptions of two colors and one flavor in the low energy. 
 
We now impose two more restrictions on this potential to place conditions on the parameters. The first is that the potential does not develop any local minima so that the inflaton rolls smoothly towards the global minimum: $V'(\phi) < 0$ for $\phi < \phi_{\rm min}$. The second is that for some field value $\phi \sim \phi_0 < M$ the potential develops an inflection point: $V''(\phi) = 0$. Applying both of these constraints we find 
\beqa
\label{eq:condition-b}
\renewcommand{\arraystretch}{1.5}
\begin{array}{cl}
b < 16 & \hspace{1cm} (V'(\phi) < 0 ~ \mbox{for}~ \phi < \phi_{\rm min})  ~,\\
b \geq 144/25 \approx 5.7&  \hspace{1cm} (\mbox{existence of inflection points for}~\phi< M) ~,
\end{array} 
\eeqa
showing the finite range for $b$. If these conditions are satisfied, there are generically two inflection points that both occur at field values $\phi_{\rm inf} < \phi_0$. 

The field point of the global minimum of the potential $\phi_{\rm min}$ can be expressed in terms of the model parameters as
\beqa
\ln\left({\frac{\phi_{\rm min}}{M}}\right) = -\frac{1}{4}\,+\, \frac{4\,b \ln{(M/\phi_0)} + \sqrt{  (c-b)(16+c-b) + 16\,b\,c\,\ln^2{(M/\phi_0)}  }  }{4\, ( c - b)} ~,
\label{eq:phimin}
\eeqa
which implies that we always have the minimum at $\phi_{\rm min}>M$. In the limit of $M \gg \phi_0$, we also have a hierarchical separation of the scales $\phi_{\rm min}$ and $M$, given by $\phi_{\rm min}/M \approx (M/\phi_0)^{(b+\sqrt{bc})/(c-b)}$. The formula of $V_0$ can be derived by requiring $V(\phi_{\rm min}) = 0$. In the large $c$ limit where the running above the threshold is very fast, $V_0 \approx M^4[1+b \ln^2(M/\phi_0)]/4$, and this gives a lower bound on $V_0$.

We now use the measured parameters of inflation to constrain the parameters of the inflation potential Eq.~\eqref{eq:pheno}. We first consider the simple case where the pivot scale of inflation, $\phi_{\rm i}$ is equal to the reference scale $\phi_0$. Using the fact that the potential is very flat around $\phi_{\rm i}$, we can approximate $V(\phi_{\rm i}) \approx a V_0$. In that case we have
\begin{eqnarray}
\epsilon_{\rm v} \approx \dfrac{\phi_0^6}{2\,V_0^2}~, \qquad \eta_{\rm v} \approx -\dfrac{(6+b)\phi_0^2}{2\,V_0}~,
\qquad \xi^2_{\rm v} \approx \dfrac{3(4+3b)\phi_0^4}{2\,V_0^2} ~, \qquad A_s \approx \dfrac{a\,V_0^3}{12\pi^2\,\phi_0^6} ~. 
\end{eqnarray}
Assuming that $\epsilon_{\rm v} \ll \eta_{\rm v}$ we can invert these expressions to get the parameters of the potential in terms of the observed values. Solving the equations for $\eta_{\rm v}$ and $A_s$, the parameters $a$ and $V_0$ are given by
\beq
a \approx \frac{12\pi^2 \,A_s\, (1-n_s)^3}{(6+b)^3} ~,  \qquad \qquad
V_0 \approx \frac{(6+b)\,\phi_0^2}{1-n_s} ~.
\label{eq:aVsimple}
\eeq
From this we see that we must have $a$ to be very small, $a \sim 10^{-15}$ mostly because the observed amplitude of scalar perturbations is very small. The overall scale of the inflaton potential is small in most models for this reason. 

From the expressions in Eq.~\eqref{eq:aVsimple} as well as our condition that the cosmological constant nearly vanishes at the global minimum described below Eq.~\eqref{eq:phimin}, we now have a consistency condition on $V_0$ which in turn can be translated into an upper bound on $M$ given parametrically by $M \lesssim \sqrt{M_{\rm Pl} \, \phi_0}\,$. For the inflation scale near the Planck scale, this means that $M$ must be only a factor of a few larger than $\phi_0$, but for low scale inflation there is significantly more available parameter space.

We can now use the above relations to predict the remaining observables
\begin{eqnarray}
r &\approx& \frac{8(1-n_s)^2\, \phi_0^2}{(6+b)^2 } \label{eq:rex} ~, \\
\alpha_s &\approx& -\frac{3\,(4+3 \, b)\, (1-n_s)^2}{(6+b)^2}  ~, \\
N_e &\approx& \frac{f(b)}{1-n_s} ~,
\end{eqnarray}
where $f(b)$ is a function that can be written in terms of exponential integrals and is independent of $\phi_0$. This function formally does depend on $\phi_{\rm e}$, the point where inflation ends, but the integral is dominated by the region around $\phi_{\rm i}$ so $f(b)$ is independent of $\phi_{\rm e}$ to very high precision. 
For the range of $144/25 < b < 16$, one has $2.49 < f(b) < 2.96$. Using the observed value of $n_s$ in Eq.~\eqref{eq:exp-ns}, we obtain the number of e-foldings to be 
\beqa
80\pm 10 < N_e < 96 \pm 12 ~,    \qquad :  ~\qquad \phi_{\rm i}=\phi_0 ~,
\label{eq:Neex}
\eeqa
where the $\pm$ error is from varying $n_s$ within its $1\sigma$ measurement, and upper and lower bounds are from varying $b$. 
This is only a mild improvement from the original CW potential and higher than one would get in standard cosmology, although these values can be accommodated with a non-minimal early universe evolution after inflation. 
For the formula of $\alpha_s$, we have assumed that $\xi^2_{\rm v} \gg \eta_{\rm v}\epsilon_{\rm v} \gg \epsilon_{\rm v}^2$, which is valid as long as $\phi_0 \lesssim M_{\rm Pl}$. Note that $\alpha_s$ is negative definite for this choice of parameter space. For $144/25 < b < 16$ and taking the 1$\sigma$ range for $n_s$ from Eq.~\eqref{eq:exp-ns}, one has 
\beqa
-(4.4\pm 1.1)\times 10^{-4} < \alpha_s < - (3.1\pm0.8)\times 10^{-4} ~,    \qquad :  ~\qquad \phi_{\rm i}=\phi_0 ~.
\eeqa
For $r$, the only observable sensitive to the scale $\phi_{\rm i}=\phi_0$ here, one has $1.6\times 10^{-5} \phi_0^2 < r < 5.5\times 10^{-5} \phi_0^2$, which is too small to be detected in the near future~\cite{Ade:2018sbj,Abazajian:2019eic} even when $\phi_0$ close to the Planck scale.

As mentioned above, the inflection points occur for $\phi < \phi_0$, so the reason the above simple slice of parameter space predicts too many e-foldings is that the initial field point is too far from the inflection point. In the more general case where $\phi_{\rm i}\neq \phi_0$, we can numerically compute the potential parameters such that we fit the inflation data. As in the analytic case above, the conclusion that $r\ll 1$ and that it is the only parameter sensitive to the overall scale still holds. Therefore, as above, we can use $n_s$ and $A_s$ to set $V_0$ and $a$. We then have two remaining parameters $b$ and $\phi_{\rm i}/\phi_0$, and we compute $N_e$ and $\alpha_s$ as a function of those parameters. To compute $N_e$, we note that the integral in Eq.~\eqref{eq:ne} is dominated by the region around $\phi_{\rm i}$ so there is very little sensitivity to $\phi_{\rm e}$. This in turn means that there is very little sensitivity to the parameters $M$ and $c$ in our phenomenological potential in Eq.~\eqref{eq:pheno}. Numerically, the sensitivity to $\phi_{\rm e}$ is at the sub-percent level as long as we are not in the regime where $c-b \lesssim 1$. 

\begin{figure}[t!]
	\label{fig:nef}
	\begin{center}
		\includegraphics[width=0.6\textwidth]{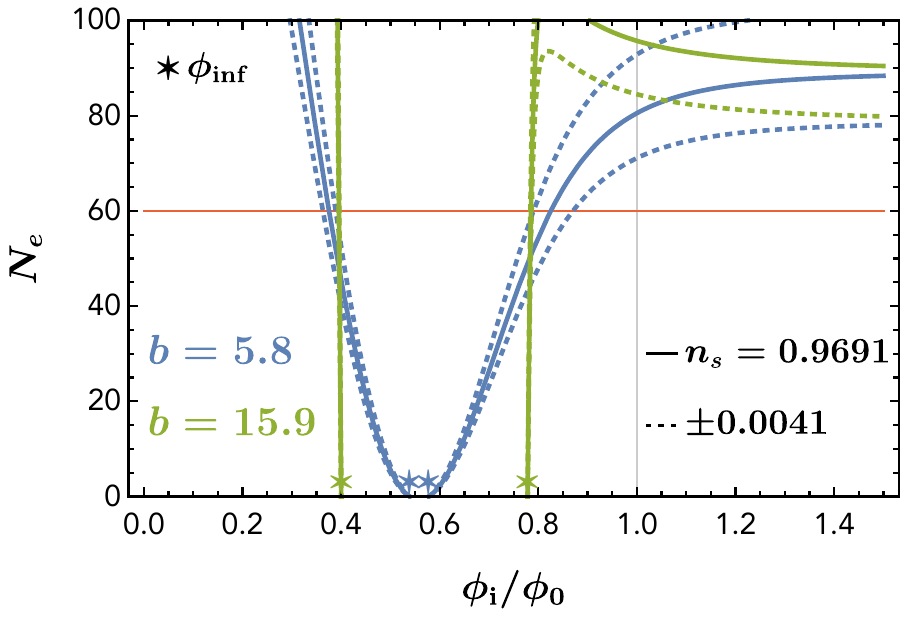} 
		\caption{Number of e-foldings $N_e$ as a function of the initial field value $\phi_{\rm i}/\phi_0$. The blue angled (green vertical) curves corresponds to $b=5.8$ ($b=15.9$). The solid lines correspond to $n_s = 0.9691$ and the dotted lines are varying $n_s$ by 0.0041, the $1\sigma$ error [see Eq.~\eqref{eq:exp-ns}]. The horizontal red line corresponds to 60 e-foldings, and the thin vertical grey line corresponds to the case of $\phi_{\rm i}/\phi_0 = 1$ analyzed semi-analytically. The locations of the inflection points in the potential are marked with stars. 
		}
	\end{center}
\end{figure}

In Fig.~\ref{fig:nef} we show the number of e-foldings as a function of $\phi_{\rm i}/\phi_0$ for $b$ near the extremes of the allowed range of Eq.~\eqref{eq:condition-b}. We see that there are two solutions for $N_e \approx 60$: $\phi_{\rm i}/\phi_0 \approx 0.8$ and $\phi_{\rm i}/\phi_0 \approx 0.4$. These correspond to the pivot scale of the inflaton field value around the two different inflection points of the potential which are also marked on the figure.\footnote{The location of the two inflection points correspond to different sets of parameters. Namely, one needs different values of $a$ and $V_0$ to fit the data depending on whether $\phi_{\rm i}/\phi_0 \approx 0.4$ or 0.8. } Note that between the inflection points there are no self-consistent solutions because in that region $V''(\phi) > 0$ and the spectrum has a blue tilt ($n_s > 1$).
 
The curve in Fig.~\ref{fig:nef} for large $b$ is approximately vertical because at $b=16$, the derivative of the potential vanishes and the integrand of Eq.~\eqref{eq:ne} diverges, so near that value of $b$, $N_e$ changes very quickly, and we must have $\phi_{\rm i}$ very close to the inflection point. On the other hand, for $b=144/25\approx 5.8$, the second derivative vanishes so for low $b$ the two inflection points are very close together. Finally, we also put a vertical grey line at $\phi_{\rm i}=\phi_0$ in Figure~\ref{fig:nef} showing that this numerical analysis reproduces the result of Eq.~\eqref{eq:Neex}. At large $\phi_{\rm i}/\phi_0$, we are far from the inflection point and the standard CW result is reproduced fitting the $n_s$ data with $\sim 90$ e-foldings.

\begin{figure}[t!]
	\label{fig:alpha}
	\begin{center}
		\includegraphics[width=0.6\textwidth]{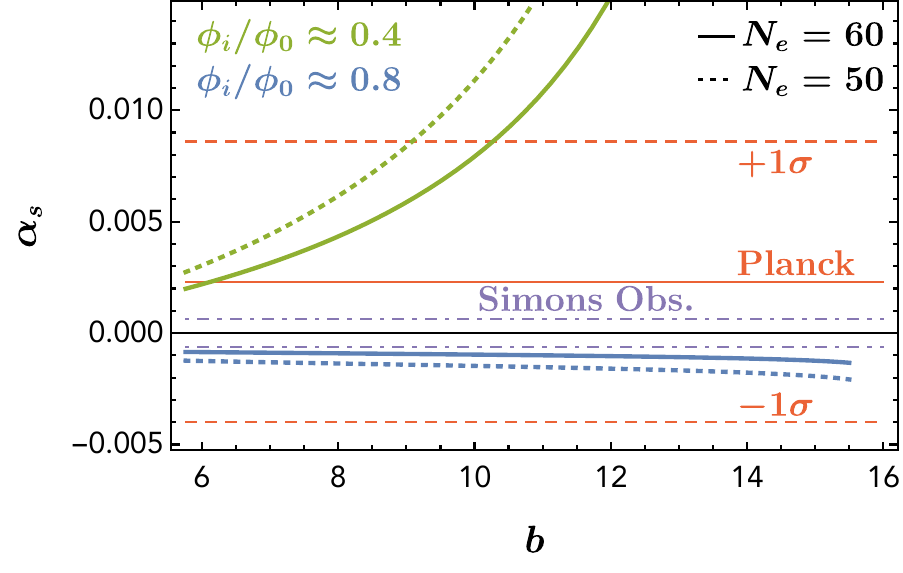}
		\caption{$\alpha_s$ vs.~$b$. The lower blue solid and dotted lines correspond to $\phi_{\rm i}/\phi_0 \approx 0.8$ and the upper solid and dotted green lines are $\phi_{\rm i}/\phi_0 \approx 0.4$. The solid lines corresponds to $N_e = 60$, while the dotted ones are $N_e = 50$. We have taken the central value of $n_s$ but the dependance is very weak. 
		The red upper horizontal solid (dashed) lines correspond to the observed central value ($\pm 1 \sigma$) for $\alpha_s$~\cite{Akrami:2018odb,Aiola:2020azj} [see Eq.~\eqref{eq:exp-alphas}], and the dot-dashed purple horizontal lines correspond to the future $1\sigma$ expected limit from the Simons Observatory assuming a central value of zero~\cite{Ade:2018sbj}.    }
	\end{center}
\end{figure}

\begin{figure}[t!]
	\label{fig:r}
	\begin{center}
	        \includegraphics[width=0.47\textwidth]{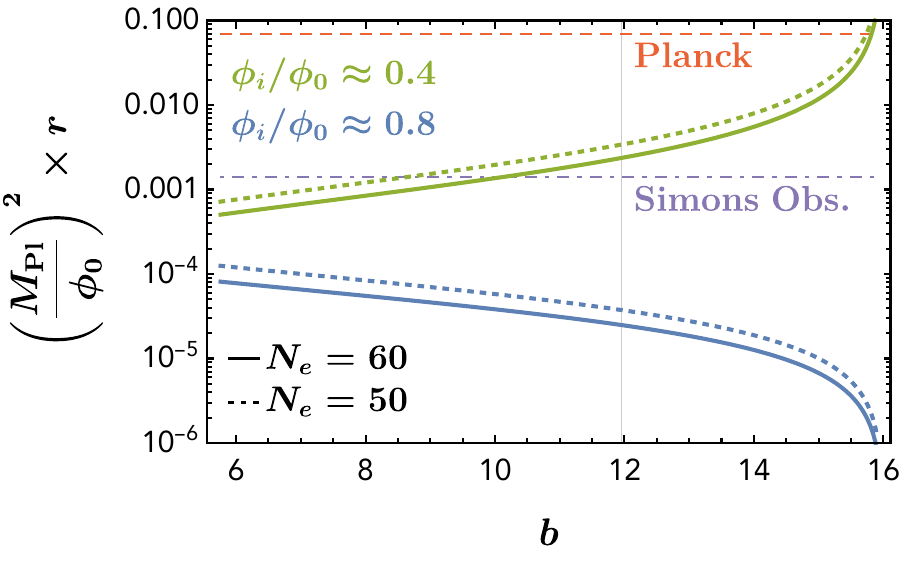} 
		 \hspace{6mm}
	        \includegraphics[width=0.47\textwidth]{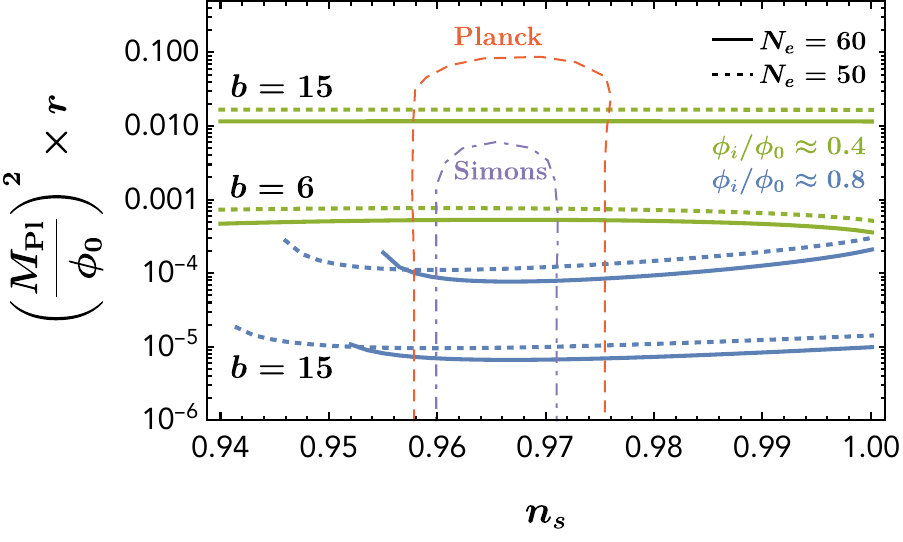} 
		\caption{\textbf{Left:} $\left(M_{\rm Pl}/\phi_0 \right)^2 \times r$ vs.~$b$ with the same lines as in Fig.~\ref{fig:alpha}.  The horizontal dashed red line is the observed upper limit [see Eq.~\eqref{eq:exp-r}] on $r$ (independent of $\phi_0$), and the horizontal purple dot-dashed line is the expected future limit from the Simons Observatory~\cite{Ade:2018sbj}. 
		The vertical thin grey line is the $2\sigma$ upper limit on $b$ for the $\phi_{\rm i}/\phi_0 \approx 0.4$ solution from the $\alpha_s$ measurement assuming $N_e = 60$.
		\textbf{Right:} $\left(M_{\rm Pl}/\phi_0 \right)^2 \times r$ vs.~$n_s$. The arcing dashed (dot-dashed) lines are present~\cite{Akrami:2018odb,Aiola:2020azj} (expected future~\cite{Ade:2018sbj}) observational limits on the $n_s-r$ plane (independent of $\phi_0$). As in the left panel, the solid lines are for $N_e = 60$, and the dotted lines are for $N_e = 50$. The bottom two and top two lines are for $b=15$, while the middle four are for $b=6$. The bottom four blue lines are for $\phi_{\rm i} / \phi_0 \approx 0.8$, while the top four green lines are for $\phi_{\rm i} / \phi_0 \approx 0.4$.  }
	\end{center}
\end{figure}

We can also use $N_e$ to determine $\phi_{\rm i}/\phi_0$ leaving us with effectively one free parameter, $b$, given the inflationary parameters. 
In Fig.~\ref{fig:alpha}, we plot $\alpha_s$ as a function of $b$ for 50 and 60 e-foldings with $n_s$ taken to its central value. The $n_s$ dependence is very weak because the measurement is so precise. As above, we have $\xi^2_{\rm v} \gg \epsilon_{\rm v} \eta_{\rm v} \gg \epsilon_{\rm v}^2$, so $\alpha_s$ is controlled by the third derivative of the potential.  For a given value $N_e$, there are two possible solutions to $\phi_{\rm i}/\phi_0$: $\phi_{\rm i}/\phi_0 \approx 0.8$ and $\phi_{\rm i}/\phi_0 \approx 0.4$.\footnote{Note also that across Figs.~\ref{fig:alpha} and~\ref{fig:r}, $\phi_{\rm i}/\phi_0$ varies around its approximate value in order to achieve the specified value for $N_e$.} The one with a larger (smaller) of $\phi_{\rm i}/\phi_0$ always has negative (positive) $\alpha_s$. This is because the $n_s$ measurement requires the second derivative of the potential to be slightly negative. Because the second derivative is going from negative to positive to negative with increasing field value, the larger (smaller) solution will be at greater (smaller) field values than the inflection point, so the third derivative of the potential will be negative (positive). From Eqs.~\eqref{eq:xi} and~\eqref{eq:alpha}, we can thus understand the signs of $\alpha_s$ for the two solutions. We also see that in the solution with smaller  $\phi_{\rm i}/\phi_0$, larger values of $b$ predict very large values of $\alpha_s$ that are inconsistent with current measurements. For the other solution with larger  $\phi_{\rm i}/\phi_0$, $\alpha_s$ is smaller than current precision but it may be probed by the next-generation CMB telescope (for instance, the Simons Observatory~\cite{Ade:2018sbj}).

Finally we analyze $r$, which depends on the additional parameter $\phi_0$. In the left panel of Fig.~\ref{fig:r} we plot $(M_{\rm Pl}/\phi_0)^2 \times r$ vs.~$b$, roughly the largest $r$ can be in the sub-Planckian regime.  At $b=16$, the first derivative of the potential vanishes at some field value, and the larger $\phi_{\rm i}/\phi_0$ solution is near the point in field space where the magnitude of the first derivative decreases as $b$ increases. The other solution thus has increasing magnitude of the first derivative as $b$ increases, and this branch does allow relatively large values of $r$ for  $\phi_0 \sim M_{\rm Pl}$. We also put a vertical line where, given 60 e-foldings, the branch with larger value of $r$ is excluded by the $\alpha_s$ measurement. This figure uses the central value of $n_s$, and we show how $r$ varies with $n_s$ in the right panel of Fig.~\ref{fig:r}. We see that within the 2$\sigma$ range for $n_s$, $r$ only changes only moderately. On the other hand, we see both panels in Fig.~\ref{fig:r} that $r$ can vary by orders of magnitude by changing $b$ or $\phi_0$. Finally we note that the branch with $\phi_{\rm i}/\phi_0 \approx 0.8$ does not admit solutions with $n_s \lesssim 1-3/N_e$, the predicted value of the original CW model discussed in Section~\ref{sec:CW}. This is why the lower blue lines on the right panel do not extend to the lowest values of $n_s$. 

\section{Attractor solutions}
\label{sec:attractor}

It is well-known that the slow-roll inflation is an attractor solution among different points in phase space~\cite{Salopek:1990jq,Vennin:2014xta}. On the other hand, ultra slow roll inflation~\cite{Kinney:2005vj,Dimopoulos:2017ged} where the slope of the potential is approximately vanishing can have different dynamics and is not necessarily an attractor solution~\cite{Pattison:2018bct}. The potential in dynamical inflection point inflation has its second derivative suppressed in order to achieve $n_s$ values consistent with observation, but the first derivative of the potential is not necessarily suppressed, and requiring $b<16$ ensures that slope is non-zero throughout the inflationary trajectory. 

In this section, we want to explicitly show that dynamical inflection point inflation also has attractor solutions and is not in the ultra slow roll regime. The inflation trajectory is the solution to the Klein-Gordon equation in a Friedmann-Robertson-Walker (FRW) background
\beqa
\ddot{\phi} + 3\,H\,\dot{\phi} + V'(\phi) = 0 ~,    \qquad \mbox{with} \quad H^2 = \frac{1}{3\,M_{\rm Pl}^2}\left[ V(\phi) + \frac{\dot{\phi}^2}{2} \right] ~,
\label{eq:KG}
\eeqa
and the slow-roll approximation that we have used throughout this work holds if
\beqa
\ddot{\phi} \ll 3\,H\,\dot{\phi} \,,    \qquad \mbox{and} \qquad  V(\phi) \gg \frac{\dot{\phi}^2}{2}  ~.
\label{eq:slow-roll}
\eeqa
These conditions ensure that the expansion of the universe is dominated by vacuum energy and that the evolution of $\phi$ is controlled by Hubble friction balancing against rolling $\phi$ down the potential. 

\begin{figure}[t!]
	\label{fig:attractor}
	\begin{center}
		\includegraphics[width=0.465\textwidth]{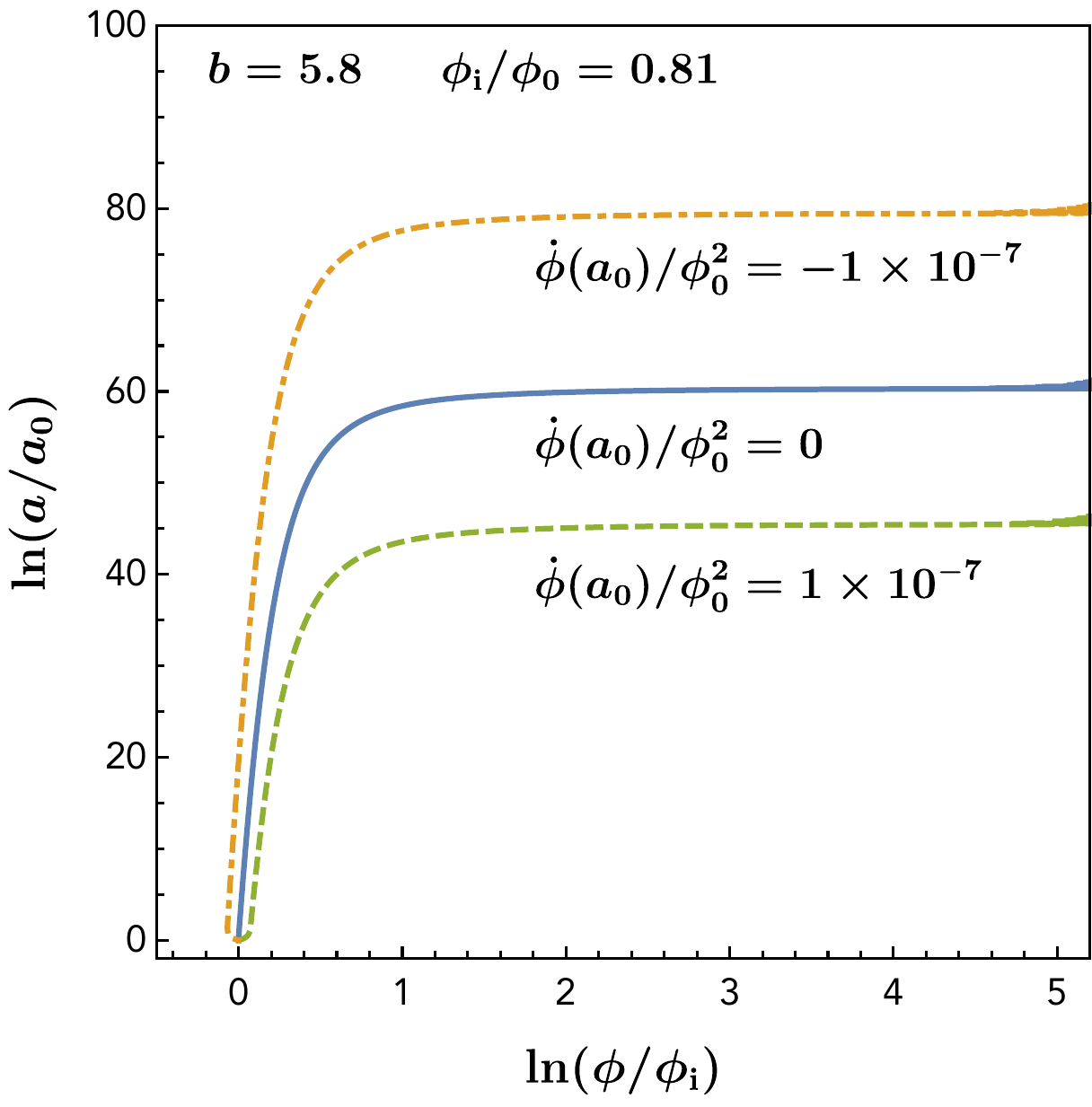}
		 \hspace{6mm}
	        \includegraphics[width=0.485\textwidth]{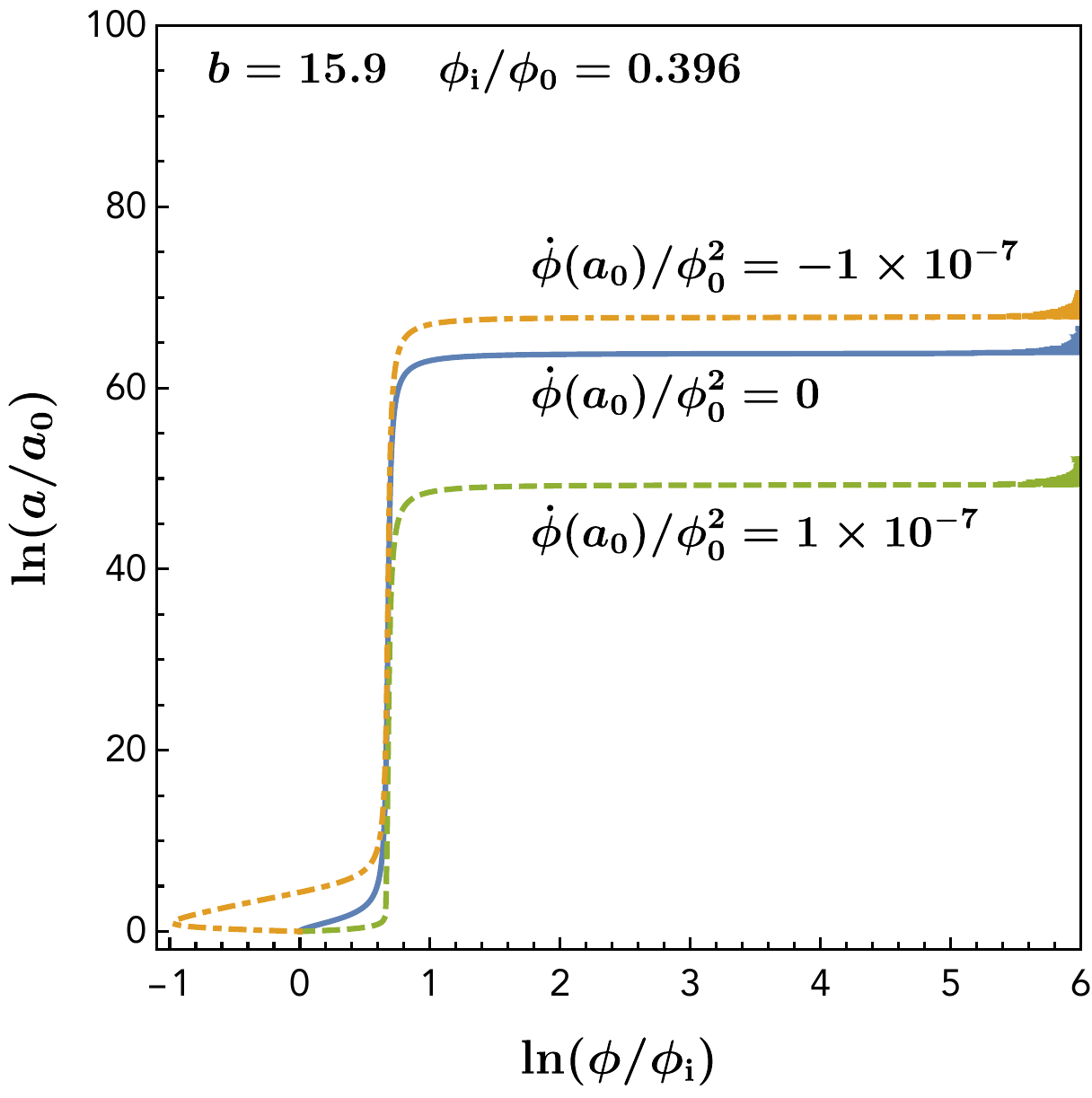} 
		\caption{\textbf{Left:} $\ln(a/a_0)$ vs.~$\ln{(\phi/\phi_{\rm i}})$ for different field velocities at the field value $\phi_{\rm i}$ or the scale factor $a = a_0$ for a benchmark model point with $b=5.8$ and $\phi_{\rm i}/\phi_0=0.81$.   
		\textbf{Right:} the same as the left panel, but for $b=15.9$ and $\phi_{\rm i}/\phi_0=0.396$.
		 }
	\end{center}
\end{figure}

We can now explore the basin of attraction for the slow-roll solution in the dynamical inflection point inflation by solving Eq.~(\ref{eq:KG}) numerically for various initial conditions. We first perturb a solution that achieves 60 e-folds, keeping the value of $\phi_{\rm i}$ the same but considering non-zero initial velocities. We set the numerical value of the initial velocity such that the conditions in Eq.~\eqref{eq:slow-roll} are not too badly violated. Of course, if the velocity is too large, then the field will roll quickly down to the minimum and slow roll is never achieved. 
We show the numerical solutions of the scale factor as a function of the field for different initial field velocities at $\phi = \phi_{\rm i}$ in Fig.~\ref{fig:attractor}. For small velocities (within the attractor phase space), the trajectories of $a$ in $\phi$ are close to each other and show attractor behaviors. 

\begin{figure}[t!]
	\label{fig:basin}
	\begin{center}
		\includegraphics[width=0.49\textwidth]{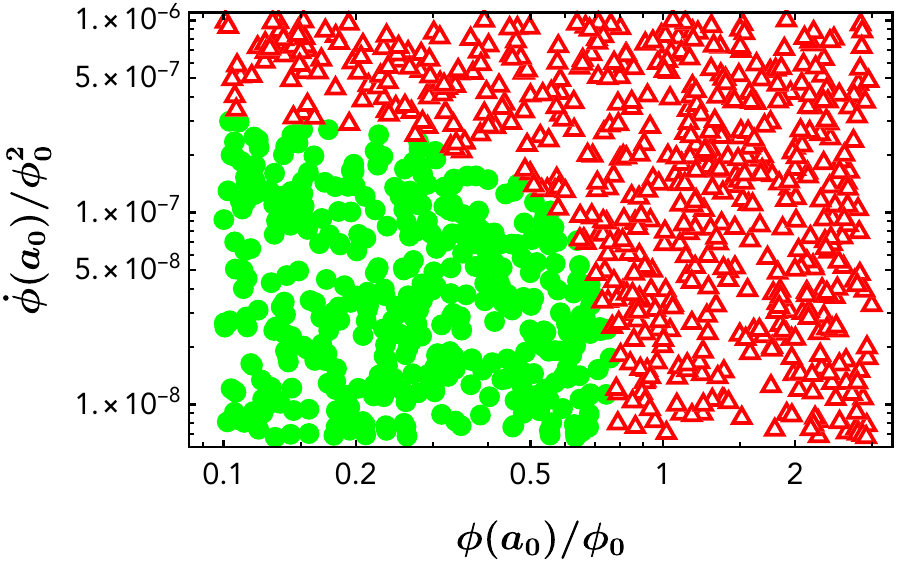}
		 \hspace{6mm}
	        \includegraphics[width=0.455\textwidth]{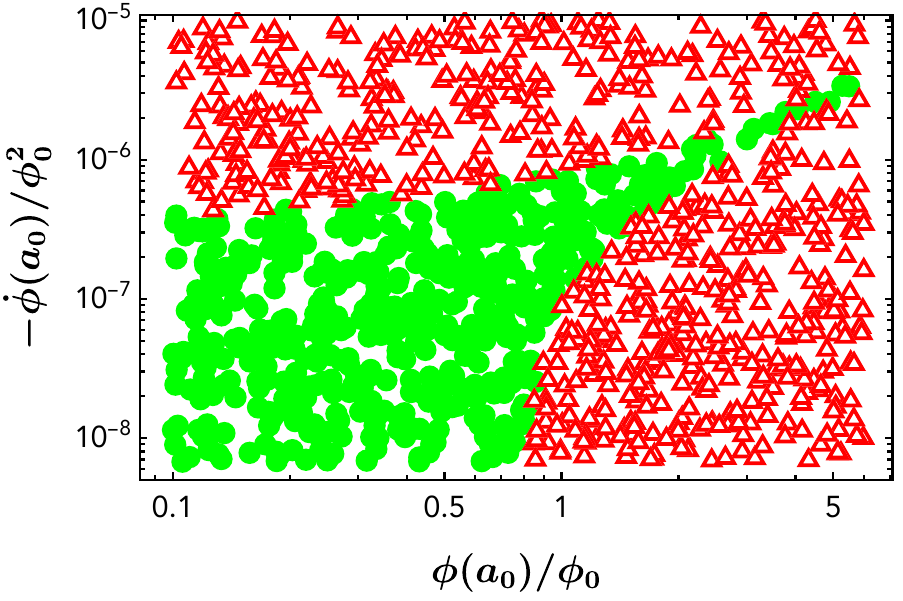} 
		\caption{The basin of attraction for the slow-roll trajectory in the $\phi(a_0)$-$\dot{\phi}(a_0)$ plane. The green filled circles are points where slow roll is reached and at least 50 e-folds of inflation occur, while the red hollow triangles are points where one of those two conditions is not satisfied. The left plot is for positive initial velocity while the right plot is for negative initial velocity. For both we take $b=10$, $c=20$, $a=10^{-15}$, and $M/\phi_0 = 2$ for the phenomenological potential in Eq.~\eqref{eq:pheno}.
		 }
	\end{center}
\end{figure}

We can also explore the phase space by noting that $\phi_{\rm i}$ is the point in field space where the cosmological scales leave the horizon, but inflation could start for much smaller values of $|\phi|$. We can thus explore the phase space of initial conditions of the inflation in the  $\phi(a_0)$-$\dot{\phi}(a_0)$ plane where $a_0$ is the scale factor when inflation begins. This is demonstrated in Fig.~\ref{fig:basin} where the green circles show points where slow roll is reached and at least 50 e-folds of inflation occur, while the red triangles are points where at least one of those conditions is violated. We use the slow roll condition from~\cite{Pattison:2018bct}:
\beq
\left| \frac{\ddot{\phi}}{ 3\,H\,\dot{\phi}} \right| < 0.1 ~. 
\label{eq:condition}
\eeq
For nearly all points where Eq.~\eqref{eq:condition} is violated, 50 e-folds of  inflation are also not achieved.

We see that very low initial velocity, at least 50 e-folds are achieved with $\phi(a_0)/\phi_0 \lesssim 0.8$, consistent with our slow-roll analysis. We also see that there is a broad basin of attraction where the slow-roll solution is reached by the time cosmological scales exit the horizon. As long as the velocity is not too large, the slow roll attractor will be reached by the time cosmological scales exit the horizon. This is true for both positive and negative initial velocities shown on the left and right panels of Fig.~\ref{fig:basin}, but for negative velocity we also see a somewhat tuned region of phase space where $\phi$ rolls up the potential before reaching the slow-roll attractor and rolling back down.

\section{Discussion and conclusions}
\label{sec:conclusion}

The inflationary paradigm is very successful at explaining the large scale features of the universe, but the search for simple and calculable models of inflation that are consistent with the precise observations from the CMB continues. Coleman-Weinberg type models are particularly attractive because they naturally produce a plateau where the potential is very flat and they can be generated dynamically by the well understood physics of radiative corrections. These models either require super-Planckian field excursions and therefore potentially suffer from large corrections due to Planck suppressed operators and possibly also quantum gravity, or they cannot be made consistent with CMB observables. 

Here we have constructed dynamical inflection point inflation which successfully gives a Coleman-Weinberg type potential that can naturally realize sub-Planckian field excursions and also be consistent with all CMB data. The key feature of these models is that the $\beta$ function of the quartic coupling of the inflaton contains two zeros that are parametrically separated. These zeros give rise to inflection points in the potential, one of which can be near the pivot scale of inflation and give the measured value of $n_s$. This can easily arise in a field theory with gauge fields and fermions when some of the fermions get a mass. The minimal model with two colors and one low energy flavor is presented in Section~\ref{sec:realization}. 

This setup has a few broad requirements on the parameters of the potential: i) the Yukawa coupling is smaller than the gauge coupling [up to some model dependent factor, see Eqs.~\eqref{eq:gysol-g}\eqref{eq:gysol} and Eq.~\eqref{eq:yboudnary}] at the scale where the fermions are integrated out, ii) the running of the gauge coupling must be faster than that of the Yukawa coupling, iii) the mass scale must not be too much larger than the scale where the $\beta$ function vanishes. Unlike other inflection point models, we \textit{do not} require a precise tuning of independent parameters in the Lagrangian. 
On the other hand, we have not addressed the ``overshoot problem" associated with small-field inflation, or how generic to have the inflation starting at a phase-space point with $\dot{\phi}_{\rm i} \approx 0$. Various possibilities to solve this problem exist including the ultra-slow roll inflation (see Ref.~\cite{Antoniadis:2020bwi} for recent discussion), but we have shown that as long as the initial velocity is not too large, there exists a broad basin of attraction for the slow-roll inflationary trajectory.


The inflationary observables of this class of models can all be computed from the phenomenological potential given in Eq.~\eqref{eq:pheno}. As far as we know, this is the first time such a potential has been studied, and this potential can be used for future studies of CMB observations. Given a value of $n_s$, $A_s$ and $N_e$, one can fully determine the potential as a function of one free parameter, $b$, which has a compact range. One can then compute the running of the scalar spectral index, $\alpha_s$, and it is predicted to be either negative with a small magnitude, or positive with a large magnitude, as shown in Fig.~\ref{fig:alpha}. The overall mass scale $\phi_0$ of the potential remains a free parameter, but as long $\phi_0 \lesssim M_{\rm Pl}$ it does not enter into any inflationary observables except $r$. This tensor to scalar ratio is expected to be below current and near future sensitivity, as show in Fig.~\ref{fig:r}.

Finally we briefly discuss other phenomenology of this model. The mass of the inflaton is parametrically given by 
\beq
m_\phi \sim \sqrt{a}\, \phi_{\rm min} = \sqrt{|\lambda_0|}\, \phi_{\rm min} \sim \frac{g_0^3}{16\pi^2} \,\phi_{\rm min} ~, \nonumber
\eeq
which is bounded from above of around $10^{-8}\,M_{\rm Pl}$ for $g_0 \sim 10^{-2}$ and $\phi_{\rm min} < M_{\rm Pl}$. The lower limit on the inflaton mass is related to how the inflaton reheats the Standard Model (SM) sector. For instance, one could introduce the operator $\lambda_{\phi h} \, \Phi^\dagger \Phi H^\dagger H$ to decay the inflaton into SM particles. To not spoil the existence of the inflection points, this coupling is required to have $\lambda_{\phi h} \ll g_0^2$. For the inflaton mass much above the Higgs boson mass, the decay width is $\Gamma(\phi \rightarrow H H^\dagger) \simeq \frac{\lambda^2_{\phi h}}{4\pi}\,m_\phi$. The reheating temperature is estimated to be $T_{\rm r} \simeq 0.2\,\sqrt{\Gamma \,M_{\rm Pl}} \simeq 10^{-2}\,\lambda_{\phi h}\sqrt{m_\phi \, M_{\rm Pl}}$. Note that when $\lambda_{\phi h} \phi_{\rm min}^2/m_\phi^2 \gtrsim 1$, preheating due to parametric resonance has to be taken into account~\cite{Kofman:1997yn}.

The hidden fermions and gauge bosons in the gauge-Yukawa model have additional model-dependent cosmological consequences. For the minimum model with $N_c = 2$ and $N_{\rm IR} = N_{\rm S} = 1$, one Dirac fermion and all gauge bosons are massive after the inflaton field sits at its minimum. The corresponding masses are $\lambda\,\phi_{\rm min}/\sqrt{2}$ and $g\,\phi_{\rm min}$, which are much heavier than the inflaton mass. The remaining fermions are massless (or massive if one introduces additional vectorlike masses for $\psi$ and $\chi$) just like the left-handed and right-handed neutrinos. One chiral fermion with the gauge interaction could have a suppressed cosmological abundance from inflaton off-shell decays. For a more general model with $N_c \ge 3$, there is generically an unbroken non-Abelian gauge symmetry after the inflaton sits at the minimum. Because of the weak interacting gauge coupling $g(\phi_0)=g_0 \sim 10^{-2}$, the confinement scale happens at an exponentially low scale [see Eq.~\eqref{eq:confine}]. So, the hidden sector mainly contains massless and weakly interacting gauge bosons. The inflaton can decay into those massless hidden gauge bosons at one-loop level (similar to the SM Higgs decays into two photons by the $W$ gauge boson loop). The decay width is estimated to be $\Gamma(\phi \rightarrow G^\mu G_\mu) \simeq g_0^4\,m_\phi^3/[4\pi (16\pi^2)^2\,\phi_{\rm min}^2]$ with $G^\mu$ representing the unbroken hidden gauge bosons. Ignoring the degrees of freedom, the ratio of the hidden sector and visible sector temperatures is
\beqa
\frac{T^{'4}}{T^4} \sim \frac{\Gamma(\phi \rightarrow G^\mu G_\mu)}{\Gamma(\phi \rightarrow H H^\dagger)} \sim \frac{g_0^4}{(16\pi^2)^2\,\lambda_{\phi h}^2}\,\frac{m_\phi^2}{\phi^2_{\rm min} }  \sim \frac{g_0^{10}}{(16\pi^2)^4\,\lambda_{\phi h}^2} ~.
\eeqa
So, for a tiny $\lambda_{\phi h}$ of the order of $g_0^5/(16\pi^2)^2$, the hidden sector contribution to the additional radiation energy of the universe could be sizable and be tested by the future CMB observations~\cite{Ade:2018sbj,Abazajian:2019eic}.

\subsubsection*{Acknowledgements}
We would like to thank Guillermo Ballesteros for useful discussion. The work of YB is supported by the U.S.~Department of Energy under the contract DE-SC-0017647. DS is supported in part by the Natural Sciences and Engineering Research Council of Canada (NSERC). We are also grateful to the The Arthur B. McDonald Canadian Astroparticle Physics Research Institute that supported YB's visit to Carleton University where this work began. YB is grateful to KITP for hospitality; this research was supported in part by the National Science Foundation under Grant No. NSF PHY-1748958. 

\appendix

\section{More general color and flavor structure}
\label{app:colourflavour}

In this appendix we briefly analyze a more general color and flavor structure of the model presented in Section~\ref{sec:realization}. We consider an $SU(N_c)$ gauge theory with gauge coupling $g$, $N_{\rm UV}$ vectorlike fundamental flavors of fermions $\psi$, $N_{\rm S}$ singlet flavors $\chi$, and a scalar inflation $\Phi$ charged under the fundamental of the gauge group. The Lagrangian is given in Eq.~\eqref{eq:lagrangian}. The $\beta$ functions for the three couplings in this theory are given by~\cite{Cheng:1973nv,Sher:1988mj}
\begin{eqnarray}
\beta_g &=& -\kappa\, g^3 \left( \frac{11}{3}N_c - \frac{1}{6}- \frac{2n_f}{3}\right) ~,
\\
\beta_{\mathbf{Y}} &=& \kappa
\left(\frac{3}{2}\,\mathbf{Y}\,\mathbf{Y}^\dagger \mathbf{Y}
+\mathbf{Y} {\rm tr}(\mathbf{Y}^\dagger \mathbf{Y})
-3\frac{N_c^2-1}{2N_c} g^2 \,\mathbf{Y} \right)~, 
\\
\beta_\lambda &=& \kappa\, \Bigg(\frac{3(N_c-1)(N_c^2 + 2N_c-2)}{4N_c^2} \,g^4 - 2\, {\rm tr}(\mathbf{Y}^\dagger \mathbf{Y}\,\mathbf{Y}^\dagger \mathbf{Y})\nonumber\\
&& \hspace{1cm} - \frac{6(N_c^2 - 1)}{N_c} \,\lambda \,g^2  + 4 \,\lambda \,{\rm tr}(\mathbf{Y}^\dagger \mathbf{Y}) + 4\,(N_c+4)\, \lambda^2 
\label{eq:betalgen}
\Bigg)~,
\end{eqnarray}
where $\kappa = (16\pi^2)^{-1}$, $n_f$ is the number of flavors kinematically accessible, and $\mathbf{Y}$ also runs over accessible flavors. As before, we search for parameter points where $\beta_\lambda = 0$. If we define 
\begin{equation}
Y^4 \equiv  {\rm tr}(\mathbf{Y}^\dagger \mathbf{Y}\,\mathbf{Y}^\dagger \mathbf{Y})
\label{eq:gengy}
\end{equation}
and work in the regime where $Y,g \gg \lambda$, then we see that $\beta_{\lambda} \approx 0$ if 
\begin{equation}
Y^4=\frac{3(N_c-1)(N_c^2 + 2N_c-2)}{8N_c^2}g^4 ~.
\label{eq:yboudnarygen}
\end{equation} 

We can now integrate the $\beta$ functions to get a leading loop approximations for the couplings as a function of $t$. 
\begin{eqnarray}
g(t) &=& g_0 - \kappa \,g_0^3 \left[ \left(\frac{11}{3}N_c - \frac{2}{3}N_{\rm IR} -\frac{1}{6}\right) t + \frac{2}{3} (N_{\rm IR}-N_{\rm UV})(t-t_M)\Theta(t-t_M) \right] ~, \\
\mathbf{Y}(t)&=& \mathbf{Y}_0 + \kappa\, t\left[\frac{3}{2}\mathbf{Y}_0\,\mathbf{Y}_0^\dagger \mathbf{Y}_0
+\mathbf{Y}_0\, {\rm tr}(\mathbf{Y}_0^\dagger \mathbf{Y}_0)- 3\,\frac{N_c^2-1}{2N_c} \,g_0^2 \,\mathbf{Y}_0 \right] ~,
\end{eqnarray}
 with $g_0$, $\mathbf{Y}_0$  being the couplings at the scale $\phi_0$, and $\mathbf{Y}_0$ and $g_0$ chosen such that Eq.~\eqref{eq:yboudnarygen} is satisfied. 
Ignoring terms proportional to $\lambda$ in Eq.~\eqref{eq:betalgen}, we can also integrate the $\beta_\lambda$ to solve for the quartic coupling as a function of scale:
 \begin{eqnarray}
 \lambda(t) &=& \lambda_0 
  - \kappa^2 t^2
\Bigg[ 
g_0^6 \times \frac{3(N_c-1)(N_c^2 + 2N_c-2)}{2N_c^2}  \times\left(\frac{11}{3}N_c - \frac{2}{3}N_{\rm IR} -\frac{1}{6}\right)+ \nonumber\\
&&4\, {\rm tr}\left( \mathbf{Y}_0^\dagger \mathbf{Y}_0\mathbf{Y}_0^\dagger \mathbf{Y}_0\right) {\rm tr}\left( \mathbf{Y}_0^\dagger \mathbf{Y}_0\right) + 
6\,{\rm tr}\left( \mathbf{Y}_0^\dagger \mathbf{Y}_0\mathbf{Y}_0^\dagger \mathbf{Y}_0\mathbf{Y}_0^\dagger \mathbf{Y}_0\right)
-6\, \frac{N_c^2-1}{N_c}\, g_0^2 \,{\rm tr}\left( \mathbf{Y}_0^\dagger \mathbf{Y}_0\mathbf{Y}_0^\dagger \mathbf{Y}_0\right)
\Bigg] \nonumber\\
&&+\frac{(N_c-1)(N_c^2 + 2N_c-2)}{N_c^2}\, \kappa^2 \,g_0^6 \,(N_{\rm UV}-N_{\rm IR})\,(t-t_M)^2\,\Theta(t-t_M) ~.
 \end{eqnarray}
If $\lambda_0 \sim \kappa^2 g_0^6\sim \kappa^2 \mathbf{Y}^6$, then ignoring the $\lambda$ terms in $\beta_\lambda$ is a good approximation. 
 
The matrix $\mathbf{Y}$ is $N_{\rm UV}\times N_{\rm S}$ for scales above $M$ and $N_{\rm IR}\times N_{\rm S}$ below $M$, and we have ignored mass terms for the singlet. In order for $\beta_\lambda$ to have two zeros as required for this inflation scenario, we need $Y^4 (M) > 3(N_c-1)(N_c^2 + 2N_c-2)g^4 (M)/({8N_c^2})$. We also need the running of $g$ to be faster than that of $\mathbf{Y}$. This can be satisfied if $ {\rm tr}(\mathbf{Y}^\dagger \mathbf{Y}\,\mathbf{Y}^\dagger \mathbf{Y})$ is dominated by light flavors. 

Unlike the minimal case, for $N_c \ge 3$ the gauge symmetry is not completely broken by the fundamental scalar vacuum expectation value and there will be a remnant non-Abelian gauge symmetry. This symmetry confines with a confinement scale given by
\begin{equation}
\Lambda = \phi_0 \,{\rm exp}\left(-\frac{3}{\kappa\,g_0^2\, (22 \, N_c - 4\, N_{\rm IR} -1)} \right).
\label{eq:confine}
\end{equation}
Because the $g_0$ required to get the right values of $a$ and $b$ in the phenomenological potential of Eq.~\eqref{eq:pheno} is small, $g_0 \lesssim 0.1$, this confinement scale is below the temperature of the universe today unless $N_c \gg 1$. Therefore we expect free massless gauge bosons with small self interactions if we are in the non-minimal model with $N_c \ge 3$.

\setlength{\bibsep}{6pt}
\bibliographystyle{JHEP}
\bibliography{fieldinflation}

\providecommand{\href}[2]{#2}\begingroup\raggedright\begin{thebibliography}{10}

\bibitem{Guth:1980zm}
A.~H. Guth, {\it {The Inflationary Universe: A Possible Solution to the Horizon
  and Flatness Problems}},  {\em Adv. Ser. Astrophys. Cosmol.} {\bf 3} (1987)
  139--148.

\bibitem{Linde:1981mu}
A.~D. Linde, {\it {A New Inflationary Universe Scenario: A Possible Solution of
  the Horizon, Flatness, Homogeneity, Isotropy and Primordial Monopole
  Problems}},  {\em Adv. Ser. Astrophys. Cosmol.} {\bf 3} (1987) 149--153.

\bibitem{Akrami:2018odb}
{\bf Planck} Collaboration, Y.~Akrami et~al., {\it {Planck 2018 results. X.
  Constraints on inflation}},  \href{http://arxiv.org/abs/1807.06211}{{\tt
  arXiv:1807.06211}}.

\bibitem{Aiola:2020azj}
S.~Aiola et~al., {\it {The Atacama Cosmology Telescope: DR4 Maps and
  Cosmological Parameters}},  \href{http://arxiv.org/abs/2007.07288}{{\tt
  arXiv:2007.07288}}.

\bibitem{Kallosh:2019jnl}
R.~Kallosh and A.~Linde, {\it {On hilltop and brane inflation after Planck}},
  {\em JCAP} {\bf 09} (2019) 030, [\href{http://arxiv.org/abs/1906.02156}{{\tt
  arXiv:1906.02156}}].

\bibitem{Guth:1982ec}
A.~H. Guth and S.~Pi, {\it {Fluctuations in the New Inflationary Universe}},
  {\em Phys. Rev. Lett.} {\bf 49} (1982) 1110--1113.

\bibitem{Shafi:2006cs}
Q.~Shafi and V.~N. Senoguz, {\it {Coleman-Weinberg potential in good agreement
  with wmap}},  {\em Phys. Rev. D} {\bf 73} (2006) 127301,
  [\href{http://arxiv.org/abs/astro-ph/0603830}{{\tt astro-ph/0603830}}].

\bibitem{Okada:2014lxa}
N.~Okada, V.~N. Senoguz, and Q.~Shafi, {\it {The Observational Status of Simple
  Inflationary Models: an Update}},  {\em Turk. J. Phys.} {\bf 40} (2016),
  no.~2 150--162, [\href{http://arxiv.org/abs/1403.6403}{{\tt
  arXiv:1403.6403}}].

\bibitem{Okada:2016ssd}
N.~Okada and D.~Raut, {\it {Inflection-point Higgs Inflation}},  {\em Phys.
  Rev. D} {\bf 95} (2017), no.~3 035035,
  [\href{http://arxiv.org/abs/1610.09362}{{\tt arXiv:1610.09362}}].

\bibitem{Urbano:2019ohp}
A.~Urbano, {\it {Inflation without gauge redundancy}},  {\em JCAP} {\bf 04}
  (2020) 040, [\href{http://arxiv.org/abs/2001.05480}{{\tt arXiv:2001.05480}}].

\bibitem{Allahverdi:2006iq}
R.~Allahverdi, K.~Enqvist, J.~Garcia-Bellido, and A.~Mazumdar, {\it {Gauge
  invariant MSSM inflaton}},  {\em Phys. Rev. Lett.} {\bf 97} (2006) 191304,
  [\href{http://arxiv.org/abs/hep-ph/0605035}{{\tt hep-ph/0605035}}].

\bibitem{Baumann:2007np}
D.~Baumann, A.~Dymarsky, I.~R. Klebanov, L.~McAllister, and P.~J. Steinhardt,
  {\it {A Delicate universe}},  {\em Phys. Rev. Lett.} {\bf 99} (2007) 141601,
  [\href{http://arxiv.org/abs/0705.3837}{{\tt arXiv:0705.3837}}].

\bibitem{Allahverdi:2008bt}
R.~Allahverdi, B.~Dutta, and A.~Mazumdar, {\it {Attraction towards an
  inflection point inflation}},  {\em Phys. Rev. D} {\bf 78} (2008) 063507,
  [\href{http://arxiv.org/abs/0806.4557}{{\tt arXiv:0806.4557}}].

\bibitem{Ballesteros:2015noa}
G.~Ballesteros and C.~Tamarit, {\it {Radiative plateau inflation}},  {\em JHEP}
  {\bf 02} (2016) 153, [\href{http://arxiv.org/abs/1510.05669}{{\tt
  arXiv:1510.05669}}].

\bibitem{Choi:2016eif}
S.-M. Choi and H.~M. Lee, {\it {Inflection point inflation and reheating}},
  {\em Eur. Phys. J. C} {\bf 76} (2016), no.~6 303,
  [\href{http://arxiv.org/abs/1601.05979}{{\tt arXiv:1601.05979}}].

\bibitem{Dimopoulos:2017xox}
K.~Dimopoulos, C.~Owen, and A.~Racioppi, {\it {Loop inflection-point
  inflation}},  {\em Astropart. Phys.} {\bf 103} (2018) 16--20,
  [\href{http://arxiv.org/abs/1706.09735}{{\tt arXiv:1706.09735}}].

\bibitem{Musoke:2017frr}
N.~Musoke and R.~Easther, {\it {Expectations for Inflationary Observables:
  Simple or Natural?}},  {\em JCAP} {\bf 12} (2017) 032,
  [\href{http://arxiv.org/abs/1709.01192}{{\tt arXiv:1709.01192}}].

\bibitem{Okada:2019yne}
N.~Okada, D.~Raut, and Q.~Shafi, {\it {Inflection-Point Inflation with Axion
  Dark Matter in light of Trans-Planckian Censorship Conjecture}},
  \href{http://arxiv.org/abs/1910.14586}{{\tt arXiv:1910.14586}}.

\bibitem{Baumann:2009ds}
D.~Baumann, {\it {Inflation}},  in {\em {Theoretical Advanced Study Institute
  in Elementary Particle Physics}: {Physics of the Large and the Small}},
  pp.~523--686, 2011.
\newblock \href{http://arxiv.org/abs/0907.5424}{{\tt arXiv:0907.5424}}.

\bibitem{Ade:2018sbj}
{\bf Simons Observatory} Collaboration, P.~Ade et~al., {\it {The Simons
  Observatory: Science goals and forecasts}},  {\em JCAP} {\bf 02} (2019) 056,
  [\href{http://arxiv.org/abs/1808.07445}{{\tt arXiv:1808.07445}}].

\bibitem{Dodelson:2003vq}
S.~Dodelson and L.~Hui, {\it {A Horizon ratio bound for inflationary
  fluctuations}},  {\em Phys. Rev. Lett.} {\bf 91} (2003) 131301,
  [\href{http://arxiv.org/abs/astro-ph/0305113}{{\tt astro-ph/0305113}}].

\bibitem{Liddle:2003as}
A.~R. Liddle and S.~M. Leach, {\it {How long before the end of inflation were
  observable perturbations produced?}},  {\em Phys. Rev. D} {\bf 68} (2003)
  103503, [\href{http://arxiv.org/abs/astro-ph/0305263}{{\tt
  astro-ph/0305263}}].

\bibitem{Coleman:1973jx}
S.~R. Coleman and E.~J. Weinberg, {\it {Radiative Corrections as the Origin of
  Spontaneous Symmetry Breaking}},  {\em Phys. Rev. D} {\bf 7} (1973)
  1888--1910.

\bibitem{Abazajian:2019eic}
K.~Abazajian et~al., {\it {CMB-S4 Science Case, Reference Design, and Project
  Plan}},  \href{http://arxiv.org/abs/1907.04473}{{\tt arXiv:1907.04473}}.

\bibitem{Salopek:1990jq}
D.~Salopek and J.~Bond, {\it {Nonlinear evolution of long wavelength metric
  fluctuations in inflationary models}},  {\em Phys. Rev. D} {\bf 42} (1990)
  3936--3962.

\bibitem{Vennin:2014xta}
V.~Vennin, {\it {Horizon-Flow off-track for Inflation}},  {\em Phys. Rev. D}
  {\bf 89} (2014), no.~8 083526, [\href{http://arxiv.org/abs/1401.2926}{{\tt
  arXiv:1401.2926}}].

\bibitem{Kinney:2005vj}
W.~H. Kinney, {\it {Horizon crossing and inflation with large eta}},  {\em
  Phys. Rev. D} {\bf 72} (2005) 023515,
  [\href{http://arxiv.org/abs/gr-qc/0503017}{{\tt gr-qc/0503017}}].

\bibitem{Dimopoulos:2017ged}
K.~Dimopoulos, {\it {Ultra slow-roll inflation demystified}},  {\em Phys. Lett.
  B} {\bf 775} (2017) 262--265, [\href{http://arxiv.org/abs/1707.05644}{{\tt
  arXiv:1707.05644}}].

\bibitem{Pattison:2018bct}
C.~Pattison, V.~Vennin, H.~Assadullahi, and D.~Wands, {\it {The attractive
  behaviour of ultra-slow-roll inflation}},  {\em JCAP} {\bf 08} (2018) 048,
  [\href{http://arxiv.org/abs/1806.09553}{{\tt arXiv:1806.09553}}].

\bibitem{Antoniadis:2020bwi}
I.~Antoniadis, A.~Chatrabhuti, H.~Isono, and S.~Sypsas, {\it {Note on initial
  conditions for small-field inflation}},
  \href{http://arxiv.org/abs/2008.02494}{{\tt arXiv:2008.02494}}.

\bibitem{Kofman:1997yn}
L.~Kofman, A.~D. Linde, and A.~A. Starobinsky, {\it {Towards the theory of
  reheating after inflation}},  {\em Phys. Rev. D} {\bf 56} (1997) 3258--3295,
  [\href{http://arxiv.org/abs/hep-ph/9704452}{{\tt hep-ph/9704452}}].

\bibitem{Cheng:1973nv}
T.~Cheng, E.~Eichten, and L.-F. Li, {\it {Higgs Phenomena in Asymptotically
  Free Gauge Theories}},  {\em Phys. Rev. D} {\bf 9} (1974) 2259.

\bibitem{Sher:1988mj}
M.~Sher, {\it {Electroweak Higgs Potentials and Vacuum Stability}},  {\em Phys.
  Rept.} {\bf 179} (1989) 273--418.

\end{thebibliography}\endgroup

\end{document}